\documentclass[aps,onecolumn,amsmath,amssymb,notitlepage,superscriptaddress]{revtex4-1}
\usepackage{graphicx}
\usepackage{dcolumn}
\usepackage{bm}
\usepackage{colonequals}
\usepackage{multirow}

\usepackage[dvipsnames]{xcolor}

\usepackage{tikz}
\usepackage[kerning=true]{microtype}
 \usepackage{ulem}

\usepackage{xr-hyper}
\definecolor{darkblue}{rgb}{0,0,0.6}
\definecolor{darkred}{rgb}{0.6,0,0}
\usepackage[colorlinks=true,urlcolor=darkblue, citecolor=darkblue, linkcolor=darkred, hyperfootnotes=false]{hyperref}

\usepackage{empheq}

\usepackage{mathrsfs}

\usepackage{enumitem}

\usepackage{dsfont}

\newcommand{\dd}{\ensuremath{\mathrm d}}

\newcommand{\dep}[2]{\ensuremath{\frac{\partial #1}{\partial #2}}}

\newcommand{\dt}[2]{\ensuremath{\frac{\dd #1}{\dd #2}}}

\def\un{\mathds{1}}

\def\D{\mathcal{D}}

\def\e{\mathrm{e}}

\newcommand{\moy}[1]{\ensuremath{\left\langle #1 \right\rangle}}

\DeclareMathOperator{\erf}{erf}
\DeclareMathOperator{\erfc}{erfc}

\begin{document}

\title{A shortcut through the macroscopic fluctuation theory: a generalised Fick law}

\author{Th\'eotim Berlioz}
\affiliation{Sorbonne Universit\'e, CNRS, Laboratoire de Physique Th\'eorique de la Mati\`ere Condens\'ee (LPTMC), 4 Place Jussieu, 75005 Paris, France}

\author{Olivier B\'enichou}
\affiliation{Sorbonne Universit\'e, CNRS, Laboratoire de Physique Th\'eorique de la Mati\`ere Condens\'ee (LPTMC), 4 Place Jussieu, 75005 Paris, France}

\author{Aur\'elien Grabsch}
\affiliation{Sorbonne Universit\'e, CNRS, Laboratoire de Physique Th\'eorique de la Mati\`ere Condens\'ee (LPTMC), 4 Place Jussieu, 75005 Paris, France}

\date{\today}

\begin{abstract}
The macroscopic fluctuation theory is a powerful tool to characterise the large scale dynamical properties of diffusive systems, both in- and out-of-equilibrium.
It relies on an action formalism in which, at large scales, the dynamics is fully determined by the minimum of the action. Within this formalism, the analysis of the statistical properties of a given observable reduces to solving the Euler-Lagrange equations with the appropriate boundary conditions. One must then compute the action at its minimum to deduce the cumulant generating function of the observable. This typically involves computing multiple integrals of cumbersome expressions.
Recently, a simple formula has been conjectured to shortcut this last step, and compute the cumulant generating function of different observables (integrated current or position of a tracer) without the need to compute any integral.
In this work, we prove this simple formula, and extend it to more general observables. 
We then illustrate the efficiency of this approach by applying it to compute the variance of a generalised current in the semi-infinite symmetric exclusion process and the joint properties of two occupation times in any diffusive system.
In the case of the integrated current, our formula can be interpreted as a generalisation of Fick's law to obtain all the cumulants of the current beyond the average value.
\end{abstract}

\maketitle

\tableofcontents

\section{Introduction}

Macroscopic fluctuation theory (MFT, not to be confused with mean field theory) is a powerful tool to study the large scale behaviour of diffusive systems. It has been developed in the 2000s by Bertini, De Sole, Gabrielli, Jona-Lasinio and Landim in a series of works~\cite{Bertini:2001,Bertini:2002,Bertini:2006,Bertini:2007}, see Ref.~\cite{Bertini:2015} for a review.
The MFT has first been developed from the study of stochastic lattice gas models like the simple exclusion process (SEP)~\cite{Bertini:2015}. Since then, it has been applied to various diffusive models in which the density is conserved, ranging from minimal models of particles or mass transfer~\cite{Bodineau:2004,Derrida:2009a,Krapivsky:2012,Krapivsky:2014,Krapivsky:2015a,Sadhu:2015,Bettelheim:2022,Bettelheim:2022a} to models of interacting Brownian particles~\cite{Grabsch:2025,Grabsch:2025c}.

The main feature of MFT is that it provides a unified description of the large scale properties of many systems in terms of only two transport coefficients: the collective diffusion coefficient $D(\rho)$, and the mobility $\sigma(\rho)$, which are both function of the local density $\rho$ of the system.
These transport coefficients are equilibrium properties of the system and encode all the microscopic details (dynamics, interaction, ...), which can then be plugged into the MFT to obtain all the dynamical and out-of-equilibrium properties of the system at large scales. The diffusion $D(\rho)$ and the mobility $\sigma(\rho)$ have been determined for various systems, ranging from minimal models of lattice gases such as the SEP~\cite{Bodineau:2004,Derrida:2009a} (see also the Table in Ref.~\cite{Rizkallah:2022} for related models), to models of Brownian particles with more realistic interactions~\cite{Lekkerkerker:1981,Cichocki:1991,Butta:1999,Felderhof:2009,Grabsch:2025,Grabsch:2025c}.

The MFT has first been successfully applied to the computation of the large deviations of the current or density in finite size diffusive system kept out-of-equilibrium by two reservoirs~\cite{Bodineau:2004,Bertini:2005a,Bodineau:2005,Sadhu:2016,Bodineau:2025,Saha:2025}. It has later been applied to study the integrated current $Q_t$ (total number of particles that cross the origin from left to right, minus those from right to left, up to time $t$) both on the infinite line~\cite{Derrida:2009a,Krapivsky:2012} and on a semi-infinite geometry~\cite{Saha:2023,Grabsch:2024d,Sharma:2024}. The MFT has also been applied to determine the distribution of the displacement $X_t$ of a tracer particle~\cite{Krapivsky:2014,Krapivsky:2015a,Sadhu:2015}.

These results rely on an action formalism, which gives the probability of observing an evolution of the macroscopic density of particles from an initial profile $\rho(x,0)$ to a given profile $\rho(x,t)$ at time $t$. Since the MFT is a macroscopic formalism, this probability is dominated by the optimal path which minimises the action. The optimal path is determined by the Euler-Lagrange equations, which yield a set of coupled nonlinear partial differential equations: the MFT equations. In theory, this formalism can be used to analyse the statistical properties of any observable, encoded in its cumulant generating function. In practice, one faces two difficulties. (i) The MFT equations cannot be solved in general, and explicit solutions are available for a few models only, including the case of the SEP~\cite{Derrida:2009a,Krapivsky:2014,Krapivsky:2015a,Sadhu:2015,Mallick:2022,Bettelheim:2022,Bettelheim:2022a,Krajenbrink:2022,Grabsch:2024}. (ii) Once the optimal evolution is determined, one must still evaluate the action along this path to obtain the cumulant generating function. This second step can be simplified using well-known thermodynamic relations (see for instance Refs.~\cite{GraTex15,CunFacViv16} for uses in the random matrix theory context), but it typically requires to compute integrals involving the optimal density profile, which can be difficult to perform in practice.

Recently, a simple formula that shortcuts step (ii) has been conjectured~\cite{Grabsch:2024b,Berlioz:2025} for any diffusive system, both for the study of the integrated current $Q_t$ or the position $X_t$ of a tracer. Here, we prove this formula and extend it to more general observables. We illustrate the efficiency of this approach by applying it to compute the variance of a generalised current in the symmetric exclusion process and the joint properties of two occupation times in any diffusive system.

\section{Macroscopic fluctuation theory}
\label{sec:MFT}

In this Section, we recall the main elements of the MFT formalism and how it can be applied to study different observables. For concreteness, we will focus on the case of the integrated current through the origin $Q_t$ in an infinite system.

\subsection{General MFT formalism}

Although the MFT can be applied to various diffusive models ranging from particle systems to mass transfer models, we will consider for concreteness a system of interacting particles at positions $\{ x_i(t) \}$, which have a stochastic dynamics. The first step is to introduce the microscopic density of particles
\begin{equation}
    \label{eq:DefDens}
    \rho_0(x,t) = \sum_{i} \delta \big( x - x_i(t) \big)
    \:,
\end{equation}
Because the number of particles is conserved, the density satisfies a conservation law
\begin{equation}
    \label{eq:ConsEq0}
    \partial_t \rho_0 + \partial_x j_0 = 0
    \:,
\end{equation}
where $j_0$ is the microscopic current, which can be defined as
\begin{equation}
    j_0(x,t) = \sum_i \dt{x_i}{t} \:
    \delta \big( x - x_i(t) \big)
    \:.
\end{equation}
The idea is now to introduce the large scale density and currents by rescaling space and time with a large parameter $\Lambda \gg 1$ (and respecting the diffusive scaling $x \sim \sqrt{t}$),
\begin{equation}
\label{eq:DefMacroFields}
    \rho(x,t) = \rho_0(\Lambda x, \Lambda^2 t)
    \:,
    \quad
    j(x,t) = \Lambda \: j_0(\Lambda x, \Lambda^2 t)
    \:,
\end{equation}
so that the macroscopic fields still satisfy the conservation relation
\begin{equation}
    \label{eq:ConsEq}
    \partial_t \rho + \partial_x j = 0
    \:.
\end{equation}
Since the microscopic dynamics is stochastic, both macroscopic fields $\rho$ and $j$ are stochastic. The main idea behind MFT is to write the current $j$ in closed form~\cite{Spohn:1983,Bertini:2015}
\begin{equation}
    \label{eq:CurrentNoise}
    j = - D(\rho) \partial_x \rho - \sqrt{\frac{\sigma(\rho)}{\Lambda}} \: \eta
    \:,
\end{equation}
where $D(\rho)$ is the collective diffusion coefficient, $\sigma(\rho)$ is the mobility and $\eta(x,t)$ is a Gaussian white noise with zero mean and unit variance,
\begin{equation}
    \label{eq:NoiseCorrel}
    \moy{\eta(x,t) \eta(x',t')} = \delta(x-x') \delta(t-t')
    \:.
\end{equation}
The constitutive equation~\eqref{eq:CurrentNoise} can be viewed as a stochastic generalisation of Fick's law. Importantly, the noise term in~\eqref{eq:CurrentNoise} is small due to the presence of the large parameter $\Lambda$, which indicates that the noise is small at large scales, even if it is not small at the microscopic scale. This is the key element that makes the MFT tractable in practice.

Indeed, Eqs.~(\ref{eq:ConsEq}-\ref{eq:NoiseCorrel}) can be used to write the probability of observing a given evolution of $\rho$ and $j$ by following a standard Martin-Siggia-Rose-Janssen-De Dominicis-Peliti
approach~\cite{Martin:1973,Janssen:1976,DeDominicis:1978}. We first write
\begin{equation}
    P[\{ \rho(x,t), j(x,t) \}] \propto
    \moy{
    \prod_{x,t}
    \delta( \partial_t \rho + \partial_x j )
    \:
    \delta\left( j +  D(\rho) \partial_x \rho + \sqrt{\frac{\sigma(\rho)}{\Lambda}} \: \eta \right)
    }_\eta
    \:,
\end{equation}
where $\moy{\cdot}_\eta$ denotes the average over the Gaussian white noise $\eta$, and the Dirac delta functions impose that Eqs.~(\ref{eq:ConsEq},\ref{eq:CurrentNoise}) are satisfied at all points in space and time, as denoted by the product over $x$ and $t$. The symbol $\propto$ indicates that the two expressions are equal up to constant prefactors which will be irrelevant. The average over the noise can be performed by first writing the second delta function as an integral,
\begin{equation}
    P[\{ \rho(x,t), j(x,t) \}] \propto
    \prod_{x,t} \delta( \partial_t \rho + \partial_x j )
    \int \D K
    \moy{
    \exp \left[
    \int \dd x \int \dd t \:
        K \left( j +  D(\rho) \partial_x \rho + \sqrt{\frac{\sigma(\rho)}{\Lambda}} \: \eta \right)
    \right]
    }_\eta
    \:,
\end{equation}
Inserting the expression of the distribution of the noise, we get
\begin{multline}
    P[\{ \rho(x,t), j(x,t) \}] 
    \propto
    \prod_{x,t}  \delta( \partial_t \rho + \partial_x j )
    \int \D K
    \int \D \eta \:
    \exp \left[
    \int \dd x \int \dd t \:
        K \left( j +  D(\rho) \partial_x \rho + \sqrt{\frac{\sigma(\rho)}{\Lambda}} \: \eta \right)
    \right]
    \\
    \times 
    \exp \left(
        - \int \dd x \int \dd t \frac{\eta^2}{2}
    \right)
    \:.
\end{multline}
The integral over $\eta$ is Gaussian, so it can be performed explicitly to yield
\begin{equation}
    P[\{ \rho(x,t), j(x,t) \}] 
    \propto
    \prod_{x,t} \delta( \partial_t \rho + \partial_x j )
    \int \D K
    \exp \left[
        \int \dd x \int \dd t \left(
            K (j + D(\rho) \partial_x \rho)
            + \frac{\sigma(\rho)}{2 \Lambda} K^2
        \right)
    \right]
    \:.
\end{equation}
The remaining integral is also Gaussian, hence we obtain,
\begin{equation}
    \label{eq:JointProbRhoJ}
    P[\{ \rho(x,t), j(x,t) \}] 
    \propto
    \prod_{x,t}\delta( \partial_t \rho + \partial_x j )
    \exp \left[
        - \Lambda \int \dd x \int \dd t
        \frac{(j + D(\rho) \partial_x \rho)^2}{2 \sigma(\rho)}
    \right]
    \:.
\end{equation}
This expression, which quantifies the probability of observing a given evolution of both $\rho$ and $j$ is a the heart of the MFT approach. Note that in the literature, Eq.~\eqref{eq:JointProbRhoJ} is often taken as the starting point. We have chosen to detail the derivation from~\eqref{eq:CurrentNoise} to stress that the Jacobian of the change of function from the noise $\eta$ to the current $j$ will be irrelevant in the following.

In the following, we will be mostly interested in observables that can be deduced from the density $\rho$, so we can integrate~\eqref{eq:JointProbRhoJ} over $j$ to obtain the marginal distribution of $\rho$. For this, it is convenient to write the last delta function as an integral,
\begin{align}
    P[\{ \rho(x,t) \}]
    &\propto
    \int \D j \: P[\{ \rho(x,t), j(x,t) \}] 
    \nonumber
    \\
    &\propto 
    \int \D H \int \D j \: \exp \left[
    -\Lambda \int \dd x \int \dd t
    \left(
    H (\partial_t \rho + \partial_x j)
    + \frac{(j + D(\rho) \partial_x \rho)^2}{2 \sigma(\rho)}
    \right)
    \right]
    \nonumber
    \\
    & \propto
    \int \D H \: \exp \left(- \Lambda \: S[\rho,H] \right)
    \:,
    \label{eq:DistrRhoMFT}
\end{align}
where we have introduced the MFT action~\cite{Bertini:2015},
\begin{equation}
    \label{eq:MFTaction}
    S[\rho,H] = \int \dd x \int \dd t \left( H \partial_t \rho + D(\rho) \partial_x \rho \partial_x H
    - \frac{\sigma(\rho)}{2} (\partial_x H)^2
    \right)
    + \int \left(
    H(+\infty,t) j(+\infty,t)
    - H(-\infty,t) j(-\infty,t)
    \right)  \dd t
    \:,
\end{equation}
where the second term comes from an integration by parts and $j(\pm \infty,t)$ are specified by boundary conditions that depend on the specific problem under consideration.
The form~\eqref{eq:DistrRhoMFT} of the probability with the action~\eqref{eq:MFTaction} will be our starting point to study different observables.

\subsection{Application to the study of the integrated current}

We consider the integrated current through the origin, defined from the microscopic system as the number of particles crossing the origin (with positive sign from left to right and negative sign from right to left), i.e.,
\begin{equation}
\label{eq:DefMicroQt}
    Q_T = \int_0^T j_0(0,t) \dd t
    = \int_0^\infty [\rho_0(x,T) - \rho_0(x,0)] \dd x
    \:,
\end{equation}
where we have assumed than the current vanishes at infinity (we will lift this hypothesis in Section~\ref{sec:WeakDrive} below).
In this equation, the last equality comes from the conservation law~\eqref{eq:ConsEq0}, that here states that the total number of particles that cross the origin is equal to the variation of the number of particles on the positive axis. Since we are interested in the long time behaviour of this quantity, we choose the rescaling parameter $\Lambda = \sqrt{T}$, so that~\eqref{eq:DefMicroQt} can be expressed as
\begin{equation}
\label{eq:DefMacroQt}
    Q_T = \sqrt{T} \int_0^\infty [\rho(x,1) - \rho(x,0)] \dd x
    \equiv \sqrt{T} Q[\rho]
\end{equation}
in terms of the macroscopic density field~\eqref{eq:DefMacroFields}. Since the integrated current $Q_T$~\eqref{eq:DefMacroQt} is expressed in terms of the macroscopic density $\rho$ only, its statistical properties are fully determined by those of the density, encoded in~\eqref{eq:DistrRhoMFT}. Therefore, we can express the moment generating function of $Q_T$ as
\begin{equation}
    \label{eq:CumulQt0}
    \moy{ \moy{\e^{\lambda Q_T}}_{\mathrm{e}} }_{\mathrm{i}}
    =
    \int \D H \int \D \rho \: \e^{ \lambda \sqrt{T}Q[\rho]} 
    P[\{ \rho(x,t) \}] P_0[\rho(x,0)]
    \:,
\end{equation}
where $\moy{\cdot}_{\mathrm{e}}$ denotes the averaging over the time evolution with~\eqref{eq:DistrRhoMFT} and $\moy{\cdot}_{\mathrm{i}}$ denotes the averaging over the initial density profile $\rho(x,0)$ with probability $P_0[\rho(x,0)]$. For simplicity, we will consider that the system is initially at equilibrium, around a mean density $\bar\rho$. Therefore, the distribution of the (macroscopic) initial density profile takes the form~\cite{Derrida:2009a}
\begin{equation}
    \label{eq:DistInit}
    P_0[\rho(x,0)] \propto \e^{-\sqrt{T} F[\rho(x,0)]}
    \:,
    \quad
    F[\rho(x,0)] = \int_{-\infty}^\infty \dd x \int_{\bar\rho}^{\rho(x,0)}
    \dd r (\rho(x,0) - r )
    \frac{2 D(r)}{\sigma(r)}
    \:,
\end{equation}
where the factor $\sqrt{T}$ in the exponential comes from the Jacobian of the change of scale~\eqref{eq:DefMacroFields} in the spatial integral~\eqref{eq:DistInit}, with $\Lambda = \sqrt{T}$.
Note that since we consider a system at equilibrium, there is no net current at infinity, $j(\pm \infty,0)$, so only the first term in the action~\eqref{eq:MFTaction} remains.
Inserting both~\eqref{eq:DistrRhoMFT} and~\eqref{eq:DistInit} into the moment generating function, we get
\begin{equation}
    \label{eq:CumulQt1}
    \moy{ \moy{\e^{\lambda Q_T}}_{\mathrm{e}} }_{\mathrm{i}}
    \propto
    \int \D H \int \D \rho \: \e^{ -\sqrt{T}( S[\rho,H] + F[\rho(x,0)] - \lambda Q[\rho])}
    \:.
\end{equation}
In the long time limit $T \to \infty$, the integrals are dominated by the optimal path $(q,p)$ of $(\rho,H)$, which is determined by the MFT equations~\cite{Derrida:2009a}
\begin{align}
    \label{eq:MFTbulk}
    \partial_t q 
    &= \partial_x [D(q) \partial_x q - \sigma(q) \partial_x p]
    \:,
    &
    \partial_t p &= - D(q) \partial_x^2 p - \frac{\sigma'(q)}{2} (\partial_x p)^2
    \:,
    \\
    \label{eq:MFTbound}
    p(x,0) &= \lambda \Theta(x) + \int_{\bar\rho}^{q(x,0)} \frac{2D(r)}{\sigma(r)} \dd r
    \:,
    &
    p(x,1) &= \lambda \Theta(x)
    \:,
\end{align}
where $\Theta$ is the Heaviside step function.
The bulk equations~\eqref{eq:MFTbulk} arise from the form of the MFT action~\eqref{eq:MFTaction}, while the initial and final conditions~\eqref{eq:MFTbound} come from the specific form of the observable~\eqref{eq:DefMacroQt} and of the initial distribution~\eqref{eq:DistInit}.
Since these boundary conditions will play an important role below, let us briefly sketch how they are derived. When looking at small variations $(\delta \rho, \delta H)$ of $S+F-\lambda Q$ around the optimal path $(q,p)$, we obtain in the double integral over space and time a term of the form $p \partial_t \delta \rho$. Performing an integration by parts in time, this yields a boundary term at $t=1$ and $t=0$ of the form $\int ( p \delta \rho \big|_{t=1} - p \delta \rho \big|_{t=0} ) \dd x$. Regrouping these boundary terms with those coming from $F$ and $Q$, we get
\begin{equation}
    \int \dd x \left\lbrace
    \delta \rho(x,1) \left[ p(x,1) - \lambda \Theta(x) \right]
    + \delta \rho(x,0) \left[ -p(x,0) + \lambda \Theta(x) + \frac{\delta F}{\delta \rho(x,0)} \right]
    \right\rbrace
     = 0
     \:.
\end{equation}
Since this should hold for any variation $\delta \rho$, we deduce the boundary conditions~\eqref{eq:MFTbound} from the expression of the free energy~\eqref{eq:DistInit}.

Finally, evaluating the integral in the moment generating function~\eqref{eq:CumulQt1} at the optimal path yields
\begin{equation}
    \label{eq:CumulQtFinal}
    \ln \moy{ \moy{\e^{\lambda Q_T}}_{\mathrm{e}} }_{\mathrm{i}}
    \underset{T \to \infty}{\simeq}
    \sqrt{T} \: \hat\psi(\lambda)
    \:,
    \quad
    \hat\psi(\lambda) = \lambda Q[q] - S[q,p] - F[q(x,0)]
    \:.
\end{equation}
All the long time behaviour of the cumulants of $Q_T$ is thus encoded in $\hat\psi$, which results from rather complicated integrals such as in the action~\eqref{eq:MFTaction}, which involves a double integral over space and time.
The determination of the cumulant generating function $\hat\psi$ thus implies two technical and potentially difficult steps:
\begin{enumerate}[label={\textbf{(Step~\arabic*)}},nosep,leftmargin=*]
    \item solve the MFT equations~(\ref{eq:MFTbulk},\ref{eq:MFTbound}) for $(q,p)$;
    \label{i:solveMFT}
    \item compute $\hat\psi$ from $(q,p)$ using~\eqref{eq:CumulQtFinal}.
    \label{i:computePsi}
\end{enumerate}
The second step can be simplified by using that $(q,p)$ is the optimal path, and thus
\begin{align}
    \dt{\hat\psi}{\lambda}
    &= \int \dd x \int \dd t \Bigg[
    \dt{q(x,t)}{\lambda} 
    \underbrace{\frac{\delta}{\delta q(x,t)} \left( \lambda Q - S -F \right)}_{=0}
    + \dt{p(x,t)}{\lambda} 
    \underbrace{\frac{\delta}{\delta p(x,t)} \left( \lambda Q - S -F \right)}_{=0}
    \Bigg]
    + Q[q]
    \nonumber
    \\
    &= Q[q]
    \label{eq:DerCumulQ}
    \:.
\end{align}
This is already an important simplification, which gets rid of the integral over $t$, but replaces it by an integral over $\lambda$ to recover $\hat\psi$.
Our aim in this article is to provide a shortcut by reducing~\ref{i:computePsi} to a simple calculation which no longer involves \textit{any} integral over $q$ or $p$, at least for observables that involve a single point in space, like $Q_t$ which measures the current through the origin only.

\section{A shortcut for the cumulant generating function of the current}
\label{sec:shortcutQt}

Recently, it has been conjectured that the cumulant generating function of the integrated current $Q_T$~\eqref{eq:CumulQtFinal} can be computed as~\cite{Grabsch:2024b}
\begin{equation}
    \label{eq:ConjPsiQt}
    \hat\psi(\lambda) = - 2 \partial_x \mu(q(x,1)) \big|_{x=0}
    \int_{q(0^-,1)}^{q(0^+,1)} D(r) \dd r
    \:,
\end{equation}
where $\mu(\rho)$ is the chemical potential that can be defined by
\begin{equation}
    \label{eq:DefChemPot}
    \mu'(\rho) = \frac{2 D(\rho)}{\sigma(\rho)}
    \:.
\end{equation}
Relation~\eqref{eq:ConjPsiQt} represents a huge simplification compared to both~\eqref{eq:CumulQtFinal} and~\eqref{eq:DerCumulQ} because (i) it only involves the solution of the MFT equation $q(x,1)$ at the final time $t=1$; (ii) the optimal profile $q(x,1)$ is involved only through its value on each side of the origin $q(0^\pm,1)$ and its derivatives at the origin $\partial_x q(0^\pm,1)$ and not through an integral over all values $q(x,1)$; (iii) it is expressed in terms of ``physical'' quantities only: the chemical potential $\mu(\rho)$ and the antiderivative of the diffusion coefficient $D(\rho)$.\\

We now prove that the conjecture~\eqref{eq:ConjPsiQt} is indeed correct. The proof relies on two conservation laws which result from the Hamiltonian structure of the MFT equations~(\ref{eq:MFTbulk}). Although we will provide alternative (and shorter) proofs in Section~\ref{sec:Extensions} below, we give here a detailed proof which sheds light on the main ingredients needed for the shortcut to work.

\subsection{Conservation laws from the Hamiltonian structure of the MFT equations}

We make use of the Hamiltonian structure of the MFT action~\eqref{eq:MFTaction}, which can be written in terms of a Hamiltonian density $\mathcal{H}(q,p)$,
\begin{equation}
    \label{eq:ActionHamilton}
    S[q,p] = \int_0^1 \dd t \int_{-\infty}^\infty \big[
    p \partial_t q - \mathcal{H}(q,p)
    \big]
    \:,
    \quad
    \mathcal{H}(q,p) = \frac{\sigma(q)}{2} (\partial_x p)^2 - D(q) \partial_x q \partial_x p
    \:.
\end{equation}
The MFT equations~(\ref{eq:MFTbulk}) are in fact the Hamilton equations
\begin{equation}
  \label{eq:HamiltEqs}
  \partial_t q = \frac{\delta \boldsymbol{H}}{\delta p} 
  \:,
  \quad
  \partial_t p = -\frac{\delta \boldsymbol{H}}{\delta q}
  \:,
  \quad \text{with} \quad
  \boldsymbol{H} =  \int_{-\infty}^\infty \mathcal{H}(q,p) \dd x
  \:.
\end{equation}
We also introduce the stress energy tensor $T^{\mu\nu}$, with $\mu, \nu \in \{0,1 \}$, which has components~\cite{Landau:1975}
\begin{equation}
    \label{eq:DefStressEnerTensor}
    T^{00} = \mathcal{H}
    \:,
    \quad
    T^{01} = p \partial_x q
    \:,
    \quad
    T^{10} = D(q) \partial_x p \partial_t q
    + (D(q) \partial_x q - \sigma(q) \partial_x p) \partial_t p
    \quad
    T^{11} = - p \partial_t q - \mathcal{H}
    \:.
\end{equation}
Since the MFT action is invariant under time translations, the energy density obeys the conservation law
\begin{equation}
    \label{eq:ConsH}
    \partial_t T^{00} + \partial_x T^{10} = 0
    \quad \Rightarrow \quad
    \partial_t \mathcal{H} + \partial_x T^{10}
    = 0
    \:.
\end{equation}
In particular, by integrating over $x$, we recover the conservation of the Hamiltonian
\begin{equation}
    \label{eq:ConsHamilt}
    \dt{\boldsymbol{H}}{t}
    = \int_{-\infty}^\infty \partial_t \mathcal{H} \dd x
    =0
    \:.
\end{equation}

We obtain a conservation law for the momentum $T^{01}$ from the invariance of the action under spatial translation,
\begin{equation}
  \label{eq:ConsT01}
  \partial_t T^{01} + \partial_x T^{11} = 0
  \quad \Rightarrow \quad
  \partial_t (p \partial_x q) - \partial_x (\mathcal{H} + p \partial_t q)
  = 0
  \:.
\end{equation}
 Note that the conserved momentum $T^{01}$ is not to be confused with the MFT momentum $p$.
Adding $-\partial_x \partial_t (qp)$ on each side of this conservation equation, and using the expression of the Hamiltonian density~\eqref{eq:ActionHamilton}, we can rewrite this conservation relation as
\begin{equation}
\label{eq:Consqdxp}
    \partial_t (q \partial_x p)
    + 
    \partial_x \left[
    \frac{\sigma(q)}{2} (\partial_x p)^2 - D(q) \partial_x q \partial_x p
    - q \partial_t p
    \right]
    = 0
    \:.
\end{equation}
We can easily check from the MFT equations~(\ref{eq:MFTbulk}) that the two conservation laws~\eqref{eq:ConsH} and~\eqref{eq:Consqdxp} are satisfied. Using these relations, we can simplify the expression of the cumulant generating function~\eqref{eq:CumulQtFinal}.

\subsection{Application to the cumulant generating function}

Starting from the conservation law~\eqref{eq:Consqdxp}, multiplying it by $x$ and integrating over $x$, gives
\begin{align}
    \dt{}{t} \int_{-\infty}^\infty x q \partial_x p \dd x
    &=
    - \int_{-\infty}^\infty x \partial_x \left[
    \frac{\sigma(q)}{2} (\partial_x p)^2 - D(q) \partial_x q \partial_x p
    - q \partial_t p
    \right] \dd x
    \nonumber
    \\
    &=
    \int_{-\infty}^\infty \left[
    \frac{\sigma(q)}{2} (\partial_x p)^2 - D(q) \partial_x q \partial_x p
    - q \partial_t p
    \right] \dd x
    \:,
\end{align}
where we have performed an integration by parts and used that $\partial_x q$ and $\partial_x p$ decay sufficiently fast at infinity. Integrating this expression over $t$ yields
\begin{align}
    \int_{-\infty}^\infty x \big[q \partial_x p \big]_{t=0}^{t=1} \dd x
    &= \int_0^1 \dd t \int_{-\infty}^\infty \dd x \left[
    \frac{\sigma(q)}{2} (\partial_x p)^2 - D(q) \partial_x q \partial_x p
    - q \partial_t p
    \right]
    \nonumber
    \\
    &= 
    \int_0^1 \dd t \int_{-\infty}^\infty \dd x \left[
    \frac{\sigma(q)}{2} (\partial_x p)^2 - D(q) \partial_x q \partial_x p
    + p \partial_t q
    \right]
    - \int_{-\infty}^{\infty} \big[q p \big]_{t=0}^{t=1} \dd x
    \nonumber
    \\
    \label{eq:RelS2}
    &= S[q,p] + 2 \int_0^1 \dd t \int_{-\infty}^\infty \dd x \left[
    \frac{\sigma(q)}{2} (\partial_x p)^2 - D(q) \partial_x q \partial_x p
    \right]
    - \int_{-\infty}^{\infty} \big[q p \big]_{t=0}^{t=1} \dd x
    \:.
\end{align}
Therefore, we have obtained from the conservation law~\eqref{eq:Consqdxp} a new expression for the optimal value of the action
\begin{equation}
    \label{eq:ActionNew1}
    S[q,p] = -2 \int_0^1 \dd t \int_{-\infty}^\infty \dd x \left[
    \frac{\sigma(q)}{2} (\partial_x p)^2 - D(q) \partial_x q \partial_x p
    \right]
    + \int_{-\infty}^\infty \big[ x q \partial_x p + q p \big]_{t=0}^{t=1} \dd x
    \:.
\end{equation}
Using now the conservation of the Hamiltonian~\eqref{eq:ConsHamilt}, we can simplify the action~\eqref{eq:ActionNew1} as
\begin{align}
    S[q,p] 
    &= 
    -2 \int_0^1 \dd t \int_{-\infty}^\infty \dd x \: \mathcal{H}(q,p)
    + \int_{-\infty}^\infty  \big[x q \partial_x p + q p \big]_{t=0}^{t=1} \dd x
    \nonumber
    \\
    &= -2 \int_{-\infty}^\infty \dd x \: \mathcal{H}(q,p) \Big|_{t=1}
    + \int_{-\infty}^\infty \big[x q \partial_x p + q p \big]_{t=0}^{t=1} \dd x
    \label{eq:ActionNew2}
    \:,
\end{align}
since $\boldsymbol{H} = \int \mathcal{H}(q,p)\dd x$ is independent of
$t$.  Note that this expression of the action results from the fact
that the Hamiltonian density is a homogeneous function of degree $2$
in $\partial_x q$ and $\partial_x p$, as discussed in
Appendix~\ref{sec:AppHomogeneousHamilt}.  Inserting the expression of
the action~\eqref{eq:ActionNew2} into the cumulant generating
function~\eqref{eq:CumulQtFinal} yields
\begin{multline}
    \label{eq:CumulNew1}
    \hat\psi(\lambda) = \lambda \int_0^\infty \big[ q(x,1) - q(x,0) \big] \dd x
    + 2 \int_{-\infty}^\infty \mathcal{H}(q,p)  \Big|_{t=1} \dd x
    - \int_{-\infty}^\infty \big[x q \partial_x p + q p \big]_{t=0}^{t=1} \dd x
    \\
    - \int_{-\infty}^\infty \dd x \int_{\bar\rho}^{q(x,0)}
    \dd r (q(x,0) - r )
    \frac{2 D(r)}{\sigma(r)}
    \:.
\end{multline}
Using now the expressions of the initial and final values of $p$ from~(\ref{eq:MFTbound}) into~\eqref{eq:CumulNew1}, several terms cancel and we get
\begin{equation}
    \hat\psi(\lambda) = 
    2 \int_{-\infty}^\infty \mathcal{H}(q,p)  \Big|_{t=1} \dd x
    - \int_{-\infty}^\infty \big[x q \partial_x p \big]_{t=0}^{t=1} \dd x
    \\
    + \int_{-\infty}^\infty \dd x \int_{\bar\rho}^{q(x,0)}
    \dd r  
    \frac{2 r D(r)}{\sigma(r)}
    \:.
\end{equation}
Together with the expression of the Hamiltonian density~\eqref{eq:ActionHamilton}, we have
\begin{equation}
    \hat\psi(\lambda) = 
    \int_{-\infty}^\infty \left[
      \partial_x p- \frac{2 D(q)}{\sigma(q)} \partial_x q
    \right] \sigma(q) \partial_x p \Bigg|_{t=1} \dd x
    - \int_{-\infty}^\infty \big[x q \partial_x p \big]_{t=0}^{t=1} \dd x
    \\
    + \int_{-\infty}^\infty \dd x \int_{\bar\rho}^{q(x,0)}
    \dd r  
    \frac{2 r D(r)}{\sigma(r)}
    \:.
\end{equation}
Finally, we perform an integration by parts in the last term to get
\begin{equation}
    \label{eq:CumulNew2}
    \hat\psi(\lambda) = 
    \int_{-\infty}^\infty \left[
      \partial_x p- \frac{2 D(q)}{\sigma(q)} \partial_x q
    \right] \sigma(q) \partial_x p \Bigg|_{t=1} \dd x
    - \int_{-\infty}^\infty x q \partial_x p \big|_{t=1} \dd x
    + \int_{-\infty}^\infty x q \left[
    \partial_x p
    -\frac{2 D(q)}{\sigma(q)} \partial_x q
    \right]
    \Bigg|_{t=0} \dd x
    \:.
\end{equation}
This expression already constitutes an important simplification compared to the original form~\eqref{eq:CumulQtFinal}, because it now only involves the optimal fields $(q,p)$ at initial and final times.
Evaluating these integrals is however tricky because the solution $q$ of the MFT equations~\eqref{eq:MFTbulk} is discontinuous both at $t=0$ and $t=1$, due to the boundary conditions~\eqref{eq:MFTbound}. This can be seen as follows. First, from the boundary condition at final time~\eqref{eq:MFTbound}, $p(x,1)$ is discontinuous at the origin. The solution $p(x,t)$ obeys an antidiffusion equation~\eqref{eq:MFTbulk}, so it is smooth for all times $t<1$. The boundary condition at $t=0$~\eqref{eq:MFTbound} therefore implies that $q(x,0)$ must be discontinuous at the origin. The solution $q(x,t)$ is obtained from the diffusion equation~\eqref{eq:MFTbulk}, which smooths it, except at $t=1$ due to the singular source term $\partial_x^2 p$, which generates a discontinuity of $q(x,1)$ at $x=0$. Looking at~\eqref{eq:CumulNew2}, this implies that all terms in the integrals are singular, because they involve products of terms in $\partial_x q$ and $\partial_x p$ and thus contain contributions proportional to $\delta(x)^2$. This is in fact not the case, as we will show that several $\delta(x)$ terms cancel out by investigating the structure of the solution of the MFT equations~(\ref{eq:MFTbulk},\ref{eq:MFTbound}) at initial and final times.

\subsection{MFT solution at initial and final times}

To simplify the expression~\eqref{eq:CumulNew2}, it is convenient to introduce the new function
\begin{equation}
    \label{eq:ChangeVarR}
    r = p - \mu(q)
    \:,
\end{equation}
with $\mu$ the chemical potential defined by~\eqref{eq:DefChemPot}. With this change of function, the MFT equations~(\ref{eq:MFTbulk},\ref{eq:MFTbound}) acquire a more symmetric form,
\begin{align}
    \label{eq:MFTbulkR}
    \partial_t r 
    &= D(q) \partial_x^2 r - \frac{\sigma'(q)}{2} (\partial_x r)^2
    \:,
    &
    \partial_t p &= - D(q) \partial_x^2 p - \frac{\sigma'(q)}{2} (\partial_x p)^2
    \:,
    \\
    \label{eq:MFTboundR}
    r(x,0) &= \lambda \Theta(x) - \mu(\bar\rho)
    \:,
    &
    p(x,1) &= \lambda \Theta(x)
    \:,
\end{align}
with $q$ obtained from $p$ and $r$ via~\eqref{eq:ChangeVarR}. These equations are symmetric under the transformation $t \to 1-t$ which exchanges $p$ and $r$, due to the time-reversal symmetry which is well-known for this problem~\cite{Derrida:2009a}. The form~(\ref{eq:MFTbulkR},\ref{eq:MFTboundR}) implies that $r$ is discontinuous at $t=0$, but smooth for $t>0$ while $p$ is discontinuous at $t=1$ and smooth for $t<1$. Writing the cumulant generating function~\eqref{eq:CumulNew2} in terms of these functions yield
\begin{equation}
    \hat\psi(\lambda) = 
    \int_{-\infty}^\infty \partial_x r \sigma(q) \partial_x p \Big|_{t=1} \dd x
    - \int_{-\infty}^\infty x q \partial_x p \big|_{t=1} \dd x
    + \int_{-\infty}^\infty x q \partial_x r
    \Big|_{t=0} \dd x
    \:.
\end{equation}
In the first term, $\partial_x p(x,1) = \lambda \delta(x)$, and $r(x,1)$ is smooth, thus
\begin{equation}
    \int_{-\infty}^\infty \partial_x r \sigma(q) \partial_x p \Big|_{t=1} \dd x
    = \partial_x r(0,1) \int_{-\infty}^\infty \sigma(q) \partial_x p \Big|_{t=1} \dd x
    \:.
\end{equation}
Similarly, the second and last term vanish because they involve $x \partial_x p(x,1) = \lambda x \delta(x)$ and $x \partial_x r(x,0) = \lambda x \delta(x)$. We end up with
\begin{equation}
    \label{eq:CumulNew3}
    \hat\psi(\lambda) = 
    \partial_x r(x,1) \int_{0^-}^{0^+} \sigma(q) \partial_x p \Big|_{t=1} \dd x
    = -\partial_x \mu(q(x,1)) \Big|_{x=0}
    \int_{0^-}^{0^+}
    \sigma\left( \mu^{-1} \big( p(x,1) - r(0,1) \big) \right) \partial_x p(x,1) \dd x
    \:,
\end{equation}
where we have used that $r(x,1)$ is smooth so that $\partial_x r(0,1) = \partial_x r(0^+,1)$ which combined with the expression of $r$~\eqref{eq:ChangeVarR} and the fact that $\partial_x p(0^+,1)=0$ gives $\partial_x r(x,1) = -\partial_x \mu(q(x,1)) \Big|_{x=0}$.
Using now that
\begin{equation}
    \partial_x \int^{\mu^{-1}( p(x,1) - r(0,1) )} D(u) \dd u
    = \frac{D \left( \mu^{-1}( p(x,1) - r(0,1) ) \right)}
    {\mu'\left( \mu^{-1}( p(x,1) - r(0,1) ) \right)} \partial_x p(x,1)
    = \frac{1}{2} \sigma \left( \mu^{-1}( p(x,1) - r(0,1) ) \right) \partial_x p(x,1)
    \:,
\end{equation}
since the chemical potential is defined by~\eqref{eq:DefChemPot}. Integrating over $x$ from $0^-$ to $0^+$ and replacing $r$ by its definition~\eqref{eq:ChangeVarR} yields
\begin{equation}
    \label{eq:IntegT1}
    \int_{0^-}^{0^+}
    \sigma\left( \mu^{-1} \big( p(x,1) - r(0,1) \big) \right) \partial_x p(x,1) \dd x
    = 2 \int_{q(0^-,1)}^{q(0^+,t)} D(r) \dd r
    \:.
\end{equation}
We present an alternative calculation of this integral in Appendix~\ref{sec:AppRegulInteg}.
Finally, inserting this result into the expression of the cumulant generating function~\eqref{eq:CumulNew3}, we obtain
\begin{equation}
    \label{eq:ShortcutQtProved}
    \boxed{
    \hat\psi(\lambda)
    = -2 \partial_x \mu(q(x,1)) \Big|_{x=0}
    \int_{q(0^-,1)}^{q(0^+,1)} D(r) \dd r
    \:.
    }
\end{equation}
This is indeed the expression~\eqref{eq:ConjPsiQt} conjectured in~\cite{Grabsch:2024b}.

\subsection{Discussion}

Let us make a few comments on the expression~\eqref{eq:ShortcutQtProved} for the cumulant generating function of the integrated current.
\begin{enumerate}[label={(\roman*)},nosep,leftmargin=*]
    \item As previously stressed, the formula~\eqref{eq:ShortcutQtProved} does not required the knowledge of the full solution $q(x,t)$ of the MFT equations to determine the cumulants, but only its values at final time on each side of the discontinuity $q(0^\pm,1)$ and its derivative $\partial_x q(x,1) \big|_{x=0^\pm}$. This constitutes an important simplification compared to the original expression~\eqref{eq:CumulQtFinal}.
    \item The strength of this approach is further emphasised by the explicit solutions of the MFT equations~(\ref{eq:MFTbulk}) obtained recently for specific models (i.e. specific choices of $D(\rho)$ and $\sigma(\rho)$)~\cite{Mallick:2022,Bettelheim:2022,Bettelheim:2022a,Krajenbrink:2022,Grabsch:2024}. These solutions were obtained from the inverse scattering approach~\cite{Ablowitz:1981} which naturally yields equations satisfied by the MFT profile at initial time $q(x,0)$ and at final time $q(x,1)$, but not at arbitrary time (the solution at all times can however be expressed as a Fredholm determinant, see for instance Ref.~\cite{Krajenbrink:2022}). This makes Eq.~\eqref{eq:ShortcutQtProved} even more practical in this context.
    \item Note that it has been shown that the shortcut~\eqref{eq:ShortcutQtProved} can be completed by two boundary conditions satisfied by the profile at final time $q(x,1)$~\cite{Grabsch:2024,Berlioz:2025}. These relations can be proved from the change of functions~\eqref{eq:ChangeVarR} and the resulting evolution equations~\eqref{eq:MFTbulkR}. As discussed above, these equations imply that $r(x,1)$ is smooth. Combining with the final condition on $p$~\eqref{eq:MFTboundR}, this implies that $\mu(q(x,1)) - \lambda \Theta(x) = -r(x,1)$ is smooth. Therefore,
    \begin{equation}
        \label{eq:BoundRelMFT}
        \mu\big( q(0^+,1) \big)
        - 
        \mu\big( q(0^-,1) \big)
        = \lambda 
        \:,
        \qquad
        \partial_x \mu \big(q(x,1) \big) \Big|_{0^+}
        = \partial_x \mu \big(q(x,1) \big) \Big|_{0^-}
        \:.
    \end{equation}
    \item Importantly, expanding the profile $q(x,1)$ at first order in $\lambda$,
    \begin{equation}
        q(x,1) = q_0(x,1) + \lambda q_1(x,1) + O(\lambda^2)
        \:,
    \end{equation}
    we can combine the shortcut~\eqref{eq:ShortcutQtProved} with the first boundary condition~\eqref{eq:BoundRelMFT} at first order,
    \begin{align}
        &\mu'(q_0(0,1)) \left[ q_1(0^+,1) - q_1(0^-,1) \right] = 1
        \:,
        \\
        &\hat\psi(\lambda) = - 2 \mu'(q_0(0,1)) \partial_x q_0(x,1) \Big|_{x=0}
        D(q_0(0,1)) \left[ q_1(0^+,1) - q_1(0^-,1) \right] \lambda + O(\lambda^2)
        \:,
    \end{align}
    to obtain
    \begin{equation}
        \label{eq:psiFick}
        \hat\psi = -2\lambda D(q_0(x,1)) \partial_x q_0(x,1) \Big|_{x=0} + O(\lambda^2)
        \:.
    \end{equation}
    This relation is actually Fick's law for the mean current. Indeed, for an initial step of density $q_0(x,0) = \rho_+ \Theta(x) + \rho_- \Theta(-x)$ (see Section~\ref{sec:ShortcutGenCurrent} below), we have that $q_0(x,t) = Q_0(x / \sqrt{t})$. Hence, Fick's law gives 
    \begin{equation}
        j_0(x,t) = - D(q_0(x,t)) \partial_x q_0(x,t) \big|_{x=0} = 
        - \frac{D(Q_0(0)) Q_0'(0)}{\sqrt{t}}
        \:.
    \end{equation}
    Therefore, integrating over $t$ in $[0,1]$, we recover the factor $2$ in~\eqref{eq:psiFick}. The shortcut~\eqref{eq:ShortcutQtProved} can therefore be seen as a generalisation of Fick's law to the full distribution of the integrated current $Q_T$. Compared to Fick's law which involves the mean density of particles, the generalisation~\eqref{eq:ShortcutQtProved} involves the MFT profile at final time, which is the large scale behaviour of~\cite{Poncet:2021,Grabsch:2022}
    \begin{equation}
        \label{eq:DefGenProf}
        \frac{\moy{ \rho_0(x \sqrt{T}, T) \: \e^{\lambda Q_T} }}
        { \moy{\e^{\lambda Q_T}} }
        \underset{T \to \infty}{\simeq} q(x,1)
        \:.
    \end{equation}
    For $\lambda = 0$, it reduces to the mean density of particles, but for $\lambda \neq 0$ it contains all the correlations $\moy{ Q_T^n \rho_0(x \sqrt{T},T)}_c$ between the current $Q_T$ and the density of particles. The quantity $q(x,1)$~\eqref{eq:DefGenProf} can thus be thought as a generalisation of the mean density which controls the full distribution of $Q_T$ through the generalised Fick law~\eqref{eq:ShortcutQtProved}.
    \item The formula~\eqref{eq:ShortcutQtProved} also applied to another observable: the position $X_T$ of a tagged particle, initially placed at the origin. Indeed, a duality relation has been proved between different models~\cite{Rizkallah:2022}. It indicates that the current $Q_T$ in a model described by $D(\rho)$ and $\sigma(\rho)$ is actually identical (up to a minus sign) to the position $X_T$ of a tracer in another model, with transport coefficients
    \begin{equation}
        \label{eq:MappingDual}
        \tilde{D}(\rho) = \frac{1}{\rho^2} D \left( \frac{1}{\rho} \right)
        \:,
        \quad
        \tilde{\sigma}(\rho) = \rho \: \sigma \left( \frac{1}{\rho} \right)
        \:.
    \end{equation}
    The idea behind this duality is that a system of particles can be equivalently described by the positions of the particles or the gaps between them. The displacement $X_T$ of a given particle (labelled $0$) is thus equivalent to a ``flux of gap'' $-Q_T$ at the origin (which is the label of the particle).
    Denoting $\moy{\cdot}^{(D,\sigma)}$ an average in a system with coefficients $D(\rho)$ and $\sigma(\rho)$, we have that
    \begin{equation}
    \label{eq:CumulRelXQ}
    \ln \moy{ \moy{\e^{-\lambda X_T}}^{(\tilde{D},\tilde\sigma)}_{\mathrm{e}} }^{(\tilde{D},\tilde\sigma)}_{\mathrm{i}}
        =
        \ln \moy{ \moy{\e^{\lambda Q_T}}^{(D,\sigma)}_{\mathrm{e}} }^{(D,\sigma)}_{\mathrm{i}}
        \underset{T \to \infty}{\simeq}
        \sqrt{T} \: \hat\psi(\lambda)
    \end{equation}
    If $q(x,1)$ is the final profile in the system described by $D$ and $\sigma$, the profile at final time in the system described by $\tilde{D}$ and $\tilde\sigma$ is given by~\cite{Rizkallah:2022}
    \begin{equation}
    \label{eq:MappingDualDens}
        \tilde{q}(y(x),1) = \frac{1}{q(x,1)}
        \:,
        \quad
        y(x) = \xi + \int_0^x q(x',1) \dd x'
        \:,
        \quad
        \xi \equiv \frac{X_T}{\sqrt{T}}
        \:.
    \end{equation}
    From these relations, we can easily show that
    \begin{equation}
        \tilde{q}(\xi^\pm,1) = \frac{1}{q(0^\pm,1)}
        \:,
        \quad
        \partial_y \tilde{q}(\xi^\pm,1) = - \frac{\partial_x q(0^\pm,1)}{q(0^\pm,1)^3}
        \:.
    \end{equation}
    Together with the transformation of the transport coefficients~\eqref{eq:MappingDual}, this implies that
    \begin{equation}
        \partial_y \tilde{\mu}(\tilde{q}) \Big|_{y=\xi^\pm} 
        =
        - \partial_x \mu(q) \Big|_{x=0^\pm}
        \:,
        \quad
        \int_{\tilde{q}(\xi^-,1)}^{\tilde{q}(\xi^+,1)} \tilde{D}(r) \dd r
        = -\int_{q(0^-,1)}^{q(0^+,1)} D(r) \dd r
        \:.
    \end{equation}
    Inserting these expressions into the expression of the cumulant generating function~\eqref{eq:ShortcutQtProved}, together with the relation~\eqref{eq:CumulRelXQ}, we get
    \begin{equation}
        \frac{1}{\sqrt{T}}
        \ln \moy{ \moy{\e^{-\lambda X_T}}^{(\tilde{D},\tilde\sigma)}_{\mathrm{e}} }^{(\tilde{D},\tilde\sigma)}_{\mathrm{i}}
        \underset{T \to \infty}{\simeq}
        -2 \partial_y \tilde{\mu}(\tilde{q}) \Big|_{y=\xi^\pm} 
        \int_{\tilde{q}(\xi^-,1)}^{\tilde{q}(\xi^+,1)} \tilde{D}(r) \dd r
        \:,
    \end{equation}
    which is exactly the same relation as for the current~\eqref{eq:ShortcutQtProved}, except that now the final profile $\tilde{q}(y,1)$ is discontinuous at the position $\xi$ of the tracer.
    \item The form of the expression~\eqref{eq:ShortcutQtProved} presents similarities with the Jensen-Varadhan functional~\cite{Jensen:2000,Varadhan:2004} which controls the large deviations of the probability of observing a given density profile in the asymmetric simple exclusion process (see Ref.~\cite{Bodineau:2006} for a discussion in the physics literature). In this case, the large deviation function is fully determined by the discontinuities of the density profile (which are shocks). This is strikingly similar to our expression~\eqref{eq:ShortcutQtProved} which involves only the discontinuity of the final density profile $q(x,1)$.
\end{enumerate}

\section{Extension of the shortcut to more general situations}
\label{sec:Extensions}

Although we have derived the shortcut relation~\eqref{eq:ConjPsiQt} for the case of the integrated current, this expression can be generalised in several directions, which we now discuss.

\subsection{Weakly driven systems}
\label{sec:WeakDrive}

We first consider the case of systems submitted to a weak external driving force $F$, which scales as $F = \nu/\Lambda$, where $\Lambda$ is the rescaling parameter from the microscopic to the macroscopic scale introduced in~\eqref{eq:DefMacroFields}. This scaling of the driving force ensures that the density keeps a diffusive behaviour. In this case, a generalisation of the conjecture~\eqref{eq:ConjPsiQt} has also been proposed~\cite{Berlioz:2025}. We prove here that this generalised conjecture is also correct.

The situation of weakly driven systems can still be described within the MFT formalism, but with the current in~\eqref{eq:ConsEq} now given by
\begin{equation}
    \label{eq:CurrentNoiseDrive}
    j = - D(\rho) \partial_x \rho + \nu \sigma(\rho) - \sqrt{\frac{\sigma(\rho)}{\Lambda}} \: \eta
    \:.
\end{equation}
The procedure described in Section~\ref{sec:MFT} can be reproduced identically, to yield the new MFT action
\begin{equation}
    \label{eq:MFTactionDrive}
    S[\rho,H] = \int_{-\infty}^\infty \dd x \int_0^1 \dd t \left[
    H \partial_t \rho + D(\rho) \partial_x \rho \partial_x H
    - \nu \sigma(\rho) \partial_x H
    - \frac{\sigma(\rho)}{2} (\partial_x H)^2
    \right]
    + \nu \sigma(\bar\rho) \int_{0}^1 (H(+\infty,t) - H(-\infty,t)) \dd t
    \:,
\end{equation}
where the last term comes from the boundary conditions $j(\pm \infty,t) = \nu \sigma(\bar\rho)$.
We again consider the integrated current through the origin up to a large time $T$ and choose the rescaling parameter $\Lambda = \sqrt{T}$. We have
\begin{equation}
     Q_T = \int_0^T j_0(0,t) \dd t
     = \sqrt{T} \left( \nu \sigma(\bar\rho) + Q[\rho] \right)
     \:,
     \quad
     Q[\rho] = \int_0^\infty [\rho(x,1) - \rho(x,0)] \dd x
     \:,
\end{equation}
due to the rescaling~\eqref{eq:DefMacroFields} and the conservation relation~\eqref{eq:ConsEq}. Note that now, even if the density is constant ($\rho(x,t) = \bar\rho$), there is a mean current $\nu \sigma(\bar\rho)$ that flows through the origin due to the driving force.
The optimal evolution $(q,p)$ of $(\rho,H)$ now obey the equations
\begin{align}
    \label{eq:MFTbulkDrive}
    \partial_t q 
    &= \partial_x [D(q) \partial_x q - \sigma(q) \partial_x p - \nu \sigma(q)]
    \:,
    &
    \partial_t p &= - D(q) \partial_x^2 p - \frac{\sigma'(q)}{2} (\partial_x p)^2 - \nu \sigma'(q) \partial_x p
    \:,
    \\
    \label{eq:MFTboundDrive}
    p(x,0) &= \lambda \Theta(x) + \int_{\bar\rho}^{q(x,0)} \frac{2D(r)}{\sigma(r)} \dd r
    \:,
    &
    p(x,1) &= \lambda \Theta(x)
    \:,
\end{align}
and the cumulant generating function of $Q_T$ is given by
\begin{equation}
    \label{eq:CumulQtFinalDrive}
    \ln \moy{ \moy{\e^{\lambda Q_T}}_{\mathrm{e}} }_{\mathrm{i}}
    \underset{T \to \infty}{\simeq}
    \sqrt{T} \: \hat\psi(\lambda;\nu)
    \:,
    \quad
    \hat\psi(\lambda;\nu) = \lambda \nu \sigma(\bar\rho) + \lambda Q[q] - S[q,p] - F[q(x,0)]
    \:.
\end{equation}
Note that, since $p(+\infty,t) = \lambda$ and $p(-\infty,t) = 0$ because of the boundary conditions~\eqref{eq:MFTboundDrive} and the evolution equations~\eqref{eq:MFTbulkDrive}, the boundary terms in the action~\eqref{eq:MFTactionDrive} becomes
\begin{equation}
    \nu \sigma(\bar\rho) \int_{0}^1 (p(+\infty,t) - p(-\infty,t)) \dd t 
    = \lambda \nu \sigma(\bar\rho)
    \:.
\end{equation}
This is exactly the mean current in~\eqref{eq:CumulQtFinalDrive}, so these terms cancel out to give
\begin{equation}
  S[q,p]  - \lambda \nu \sigma(\bar\rho)
  = \int_{-\infty}^\infty \dd x \int_0^1 \dd t \left[
    p \partial_t q + D(q) \partial_x q \partial_x p
    - \nu \sigma(q) \partial_x p
    - \frac{\sigma(q)}{2} (\partial_x p)^2
  \right]
  \:.
\end{equation}
Note that in this case, the Hamiltonian density is not a homogeneous function of $\partial_x q$ and $\partial_x p$ due to the additional term $\nu \sigma(q) \partial_x p$. The method used in Section~\ref{sec:shortcutQt} can nevertheless be extended to this situation, as discussed in Appendix~\ref{sec:AppHomogeneousHamilt}. Nevertheless, we propose here an alternative derivation to the one presented above for the shortcut for $\hat\psi$ in this case, since it will allow us to present in Section~\ref{sec:UniversalShortcut} a unified description of the shortcut in all the situations considered in this article.

The first step is to reintroduce the dependence on $T$ of the action by defining
\begin{equation}
    \label{eq:DefRescQP}
    \hat{q}(x,t;T) = q \left( \frac{x}{\sqrt{T}}, \frac{t}{T} \right)
    \:,
    \quad
    \hat{p}(x,t;T) = p \left( \frac{x}{\sqrt{T}}, \frac{t}{T} \right)
    \:,
\end{equation}
so that
\begin{equation}
    \label{eq:CumulQtFinalDriveScaled}
    \ln \moy{ \moy{\e^{\lambda Q_T}}_{\mathrm{e}} }_{\mathrm{i}}
    \underset{T \to \infty}{\simeq}
    \psi(\lambda, \nu, T) 
    \:,
    \quad
    \psi(\lambda, \nu, T) 
    \equiv
    \lambda Q_T[\hat{q}] - S_T[\hat{q},\hat{p}] - F[\hat{q}(x,0)]
    \:.
\end{equation}
The functional $F[q]$ is unchanged, and still given by~\eqref{eq:DistInit}, while we have introduced
\begin{align}
    Q_T[\hat{q}] &\equiv \int_0^\infty [\hat{q}(x,T;T) - \hat{q}(x,0;T)] \dd x
    \:,
    \\
    \label{eq:ActionT}
    S_T[\hat{q},\hat{p}] &\equiv 
    \int_{-\infty}^\infty \dd x \int_0^T \dd t \left[
    \hat{p} \partial_t \hat{q} + D(\hat{q}) \partial_x \hat{q} \partial_x \hat{p}
    - \frac{\nu}{\sqrt{T}} \sigma(\hat{q}) \partial_x \hat{p}
    - \frac{\sigma(\hat{q})}{2} (\partial_x \hat{p})^2
    \right]
    \:.
\end{align}
Note that the rescaling~\eqref{eq:DefRescQP} reverses the change of scale~\eqref{eq:DefMacroFields}. This means that $\hat{q}$ and $\hat{p}$ depend on the microscopic parameters $x$ and $t$, but these are not the true microscopic fields. They are only a rescaling of the macroscopic fields $q$ and $p$ which give a coarse-grained description of the microscopic dynamics.
The new fields $(\hat{q},\hat{p})$ obey the same MFT equations~(\ref{eq:MFTbulkDrive},\ref{eq:MFTboundDrive}), except that the final condition is now at $t=T$ and $\nu$ has been replaced by the force $f = \nu/\sqrt{T}$. Importantly, these equations are also obtained from minimising $\lambda Q_T[\hat{q}] - S_T[\hat{q},\hat{p}] - F[\hat{q}(x,0)]$ with respect to $(\hat{q},\hat{p})$. A direct consequence is that
\begin{align}
    \dt{}{T} \psi \left(\lambda,\nu,T \right)
    =& {}\: \int_{-\infty}^\infty \dd x \int_0^T \dd t \Bigg[
    \dep{\hat{q}(x,t;T)}{T} 
    \underbrace{\frac{\delta}{\delta \hat{q}} \left( \lambda Q_T - S_T -F \right)}_{=0}
    + \dep{\hat{p}(x,t;T)}{T} 
    \underbrace{\frac{\delta}{\delta \hat{p}} \left( \lambda Q_T - S_T -F \right)}_{=0}
    \Bigg]
    \nonumber
    \\
    &{}\: + \partial_T \left(\lambda Q_T[\hat{q}] - S_T[\hat{q},\hat{p}] \right)
    \nonumber
    \\
    =&  {}\:
    - \int_{-\infty}^\infty \left[
    \hat{p} \partial_t \hat{q} + D(\hat{q}) \partial_x \hat{q} \partial_x \hat{p}
    - \frac{\nu}{\sqrt{T}} \sigma(\hat{q}) \partial_x \hat{p}
    - \frac{\sigma(\hat{q})}{2} (\partial_x \hat{p})^2
    \right]
    \Bigg|_{t=T}
    \dd x
    \nonumber
    \\
    & {}\:
    + \lambda \int_{0}^\infty \partial_t \hat{q}(x,t;T) \bigg|_{t=T} \dd x
    - \frac{\nu}{2 T^{3/2}} \int_{-\infty}^\infty \dd x \int_0^T \dd t \: \sigma(\hat{q}) \partial_x \hat{p}
    \:.
    \label{eq:dtPsiDriven0}
\end{align}
Similarly, we have
\begin{align}
    \dt{}{\nu} \psi \left(\lambda,\nu,T \right)
    =& {}\: \int_{-\infty}^\infty \dd x \int_0^T \dd t \Bigg[
    \dep{\hat{q}(x,t;T)}{\nu} 
    \underbrace{\frac{\delta}{\delta \hat{q}} \left( \lambda Q_T - S_T -F \right)}_{=0}
    + \dep{\hat{p}(x,t;T)}{\nu} 
    \underbrace{\frac{\delta}{\delta \hat{p}} \left( \lambda Q_T - S_T -F \right)}_{=0}
    \Bigg]
    \nonumber
    \\
    &{}\: + \partial_\nu \left(\lambda Q_T[\hat{q}] - S_T[\hat{q},\hat{p}] \right)
    \nonumber
    \\
    =& {}\: \frac{1}{\sqrt{T}} \int_{-\infty}^\infty \dd x \int_0^T \dd t \: \sigma(\hat{q}) \partial_x \hat{p}
    \:.
    \label{eq:dnuPsiDriven0}
\end{align}
By combining~\eqref{eq:dtPsiDriven0} and~\eqref{eq:dnuPsiDriven0} and using the final condition on $\hat{p}$~\eqref{eq:MFTboundDrive}, we obtain,
\begin{equation}
    \dt{}{T} \psi \left(\lambda,\nu,T \right)
    =
    -\frac{1}{2 T} \nu \dt{}{\nu} \psi \left(\lambda,\nu,T \right)
     - \int_{-\infty}^\infty \left[
    D(\hat{q}) \partial_x \hat{q} \partial_x \hat{p}
    - \frac{\nu}{\sqrt{T}} \sigma(\hat{q}) \partial_x \hat{p}
    - \frac{\sigma(\hat{q})}{2} (\partial_x \hat{p})^2
    \right]
    \Bigg|_{t=T}
    \dd x
    \:.
\end{equation}
Using that, by definition (see Eqs.~(\ref{eq:CumulQtFinalDrive},\ref{eq:CumulQtFinalDriveScaled})),
\begin{equation}
    \psi \left(\lambda,\nu,T \right) = \sqrt{T} \: \hat\psi(\lambda;\nu)
    \:,
\end{equation}
and the scalings~\eqref{eq:DefRescQP} to reintroduce $(q,p)$ instead of $(\hat{q},\hat{p})$, we get
\begin{equation}
    \label{eq:PsiDrivenNew}
    \hat\psi
    + \nu \partial_\nu \hat\psi
    =
     - 2\int_{-\infty}^\infty \left[
   \frac{2 D(q) \partial_x q}{\sigma(q)}
    - 2\nu 
    - \partial_x p
    \right]\frac{\sigma(q)}{2}  \partial_x p
    \Bigg|_{t=1}
    \dd x
    \:.
\end{equation}
Note that in the case $\nu=0$, this reduces to~\eqref{eq:CumulNew2} above, without using the conservation relation~\eqref{eq:Consqdxp} or the Hamiltonian structure of the action.

There now remains to analyse the integral in~\eqref{eq:PsiDrivenNew}. This can be done as previously by first introducing the function $r(x,t)$ as in~\eqref{eq:ChangeVarR}, which transforms the MFT equations~(\ref{eq:MFTbulkDrive},\ref{eq:MFTboundDrive}) into
\begin{align}
    \label{eq:MFTbulkRDrive}
    \partial_t r 
    &= D(q) \partial_x^2 r - \frac{\sigma'(q)}{2} (\partial_x r)^2 - \nu \sigma'(q) \partial_x r
    \:,
    &
    \partial_t p &= - D(q) \partial_x^2 p - \frac{\sigma'(q)}{2} (\partial_x p)^2 - \nu \sigma'(q) \partial_x p
    \:,
    \\
    \label{eq:MFTboundRDrive}
    r(x,0) &= \lambda \Theta(x) - \mu(\bar\rho)
    \:,
    &
    p(x,1) &= \lambda \Theta(x)
    \:.
\end{align}
Since $r$ obeys a diffusion equation with a step initial density, it becomes smooth for $t>0$, so that $\partial_x r(x,1)$ is continuous. This is the term that appears in the bracket in the integral in Eq.~\eqref{eq:PsiDrivenNew}. It is multiplied by $\partial_x p(x,1) = \lambda \delta(x)$, hence, the cumulant generating function~\eqref{eq:PsiDrivenNew} becomes
\begin{equation}
    \label{eq:PsiDrivenNew2}
    \hat\psi
    + \nu \partial_\nu \hat\psi
    = 
     - 
    \left[
   \frac{2 D(q) \partial_x q}{\sigma(q)}
    - 2\nu
    \right] \Bigg|_{x=0}
     \int_{0^-}^{0^+} \sigma \left( \mu^{-1} \big( p(x,1) - r(0,1) \big) \right)  \partial_x p(x,1)
    \dd x
    \:.
\end{equation}
We have already computed the remaining integral above, see Eq.~\eqref{eq:IntegT1}, therefore, we finally obtain from~\eqref{eq:PsiDrivenNew2},
\begin{equation}
    \label{eq:PsiShortcutDrive}
    \boxed{
    \hat\psi
    + \nu \partial_\nu \hat\psi
    =
     - 2
    \left[
   \partial_x \mu(q(x,1))
    - 2\nu
    \right] \Big|_{x=0}
     \int_{q(0^-,1)}^{q(0^+,1)} D(r) \dd r
    \:.
    }
\end{equation}
This is exactly the relation that has recently been conjectured in the case of weakly driven systems~\cite{Berlioz:2025}.

\subsection{Generalised current with a step initial density}
\label{sec:ShortcutGenCurrent}

We consider a generalisation of the integrated current through the origin $Q_T$~\eqref{eq:DefMicroQt} by considering the integrating current through a moving boundary at the microscopic position $x_t$ at time $t$. It is thus defined as
\begin{equation}
    \label{eq:DefGenIntegCurrent0}
    Q_T[x] = \int_0^T \left( j_0(x_{t},t) - \dt{x_{t}}{t} \rho_0(x_{t},t) \right) \dd t
    \:,
\end{equation}
where the second term accounts for the particles which cross the position $x_t$ due to the movement of the fictitious boundary, and not due to the movements of the particles (encoded in the first term). This class of observable allows one to study the dynamics of individual particles, since $Q_T[x] = 0$ implies that the particle that was initially at $x_0$ is now located at $x_T$ (because the order of the particles is conserved).
Note that, in the context of  exclusion processes, $Q_T[x]$ is also called a height function, since it is involved in a classical mapping onto an interface model~\cite{Imamura:2017,Imamura:2021}.
Using the microscopic conservation law~\eqref{eq:ConsEq0}, the generalised integrated current~\eqref{eq:DefGenIntegCurrent0} can be rewritten in a more convenient form,
\begin{equation}
    \label{eq:DefGenIntegCurrent1}
    Q_T[x] = \int_0^\infty \left[ 
    \rho_0(x_T + x, T) - \rho_0(x_0 + x,0)
    \right]
    \dd x
    + j_0(+\infty) T
    - \rho_0(+\infty) \: (x_T - x_0)
    \:.
\end{equation}
Since we are interested in diffusive system, we focus on the situation where $x_t$ follows a diffusive scaling $x_t \underset{t \to \infty}{\simeq} \xi \sqrt{t}$. Introducing the macroscopic fields by performing the rescaling~\eqref{eq:DefMacroFields} with $\Lambda = \sqrt{T}$, we obtain
\begin{equation}
    \label{eq:DefGenIntegCurrentMacro}
    Q_t[x] = \sqrt{T} \left[ \int_0^\infty \left[ 
    \rho(\xi + x, 1) - \rho(\xi_0 + x,0)
    \right]
    \dd x
    + j(+\infty)
    - \rho(+\infty) \: (\xi - \xi_0)
    \right]
    \:,
\end{equation}
where we have also scaled the initial position $x_0$ with the observation time $T$, as $x_0 \underset{T \to \infty}{\simeq} \xi_0 \sqrt{T}$.

Up to now, we have considered the case of an equilibrium initial density, with a uniform mean density $\bar\rho$. We will now consider the case of a step initial density, corresponding to a system initially locally at equilibrium around a mean density
\begin{equation}
    \label{eq:StepInit}
    \bar\rho(x) = \rho_+ \Theta(x) + \rho_- \Theta(-x)
    \:.
\end{equation}
In this case, $\rho(+\infty) = \rho_+$ and $j(+\infty) = \nu \sigma(\bar\rho)$. Therefore, the generalised current~\eqref{eq:DefGenIntegCurrent0} becomes, in terms of the macroscopic quantities
\begin{equation}
    \label{eq:DefGenIntegCurrentMacroFinal}
    Q_T[x] = \sqrt{T} Q[\rho; \xi, \xi_0]
    \:,
    \quad
    Q[\rho; \xi, \xi_0] \equiv 
    \int_0^\infty \left[ 
    \rho(\xi + x, 1) - \rho(\xi_0 + x,0)
    \right]
    \dd x
    + \nu \sigma(\rho_+)
    - \rho_+ \: (\xi - \xi_0)
    \:.
\end{equation}

Our goal is to show that the shortcut~\eqref{eq:ShortcutQtProved} can be extended to this observable and to the step initial condition~\eqref{eq:StepInit}. The cumulant generating function of the integrated current takes the form
\begin{equation}
    \label{eq:CumulGenCurrent}
    \hat\psi(\lambda; \nu) 
    \equiv \lim_{T \to \infty} \frac{1}{\sqrt{T}}
    \ln \moy{\moy{
    \exp \left[
    \lambda  Q_T \left[x \right]
    \right]
    }_{\mathrm{e}}}_{\mathrm{i}}
    \:.
\end{equation}
We follow the approach of Section~\ref{sec:WeakDrive} and introduce the rescaled optimal field ($\hat{q},\hat{p}$) as in~\eqref{eq:DefRescQP}. The cumulant generating function~\eqref{eq:CumulGenCurrent} takes the form
\begin{equation}
    \label{eq:ScaledPsiGenCurrent}
    \psi(\lambda; \nu,T) \equiv \sqrt{T} \hat\psi(\lambda; \nu) 
    = \lambda Q_T \left[\hat{q}; \xi, \xi_0 \right] - S_T[\hat{q},\hat{p}] - F[\hat{q}(x,0;T)]
    \:,
\end{equation}
where we have defined $S_T$ as in~\eqref{eq:ActionT} and
\begin{equation}
    Q_T[\hat{q}; \xi, \xi_0] \equiv
    \int_0^\infty \left[ 
    \hat{q}(\xi \sqrt{T} + x, T;T) - \hat{q}(\xi_0 \sqrt{T} + x,0;T)
    \right]
    \dd x
    - \rho_+ \: (\xi - \xi_0) \sqrt{T}
    \:,
\end{equation}
with $F$ now given by
\begin{equation}
    F[\hat{q}(x,0;T)] = 
    \int_{-\infty}^\infty \dd x \int_{\bar\rho(x)}^{\hat{q}(x,0;T)}
    \dd r (\hat{q}(x,0;T) - r )
    \frac{2 D(r)}{\sigma(r)}
\end{equation}
to account for the step initial density~\eqref{eq:StepInit}.
Note that the deterministic part $\nu \sigma(\rho_+)$ of the current~\eqref{eq:DefGenIntegCurrentMacroFinal} has canceled out with the boundary terms of the action~\eqref{eq:MFTactionDrive}.
The fields $\hat{q}$ and $\hat{p}$ are now the minimum of $\lambda Q_T - S_T - F$, which give the same MFT equations~\eqref{eq:MFTbulkDrive} but the new boundary conditions
\begin{equation}
    \label{eq:BoundCdtGenCurrent}
    \hat{p}(x;T,T) = \lambda \Theta \left( x - \xi \sqrt{T} \right)
    \:,
    \quad
    \hat{p}(x;0,T) = \lambda \Theta \left( x - \xi_0 \sqrt{T} \right)
    + \int_{\bar\rho(x)}^{\hat{q}(x;0,T)} \frac{2 D(r)}{\sigma(r)} \dd r
    \:.
\end{equation}
Using that $(\hat{q},\hat{p})$ minimise the action, we obtain that
\begin{align}
    \dt{}{T}  \psi(\lambda; \nu,T)
    =& {} \: \lambda \partial_T Q_T\left[\hat{q}; \xi, \xi_0 \right]
    - \partial_T S_T[\hat{q},\hat{p}]
    \nonumber 
    \\
    =& {} \:
    \lambda \int_0^\infty \dd x \left[
        \frac{1}{2 \sqrt{T}} \partial_x \left(
        \xi \hat{q}(\xi\sqrt{T} + x, T;T) - \xi_0\hat{q}(\xi_0 \sqrt{T} + x,0;T)
        \right)
        + \partial_t \hat{q}(\xi \sqrt{T} + x, t;T) \Big|_{t=T}
    \right]
    \nonumber
    \\
    &{}\:
    - \frac{1}{2 \sqrt{T}} \lambda
    \rho_+ \left( \xi - \xi_0 \right)
    - \frac{\nu}{2 T^{3/2}} \int_{-\infty}^\infty \dd x \int_0^T \dd t \: \sigma(\hat{q}) \partial_x \hat{p} 
    \nonumber
    \\
    &{}\: - \int_{-\infty}^\infty \left[
    \hat{p} \partial_t \hat{q} + D(\hat{q}) \partial_x \hat{q} \partial_x \hat{p}
    - \frac{\nu}{\sqrt{T}} \sigma(\hat{q}) \partial_x \hat{p}
    - \frac{\sigma(\hat{q})}{2} (\partial_x \hat{p})^2
    \right]
    \Bigg|_{t=T}
    \dd x
    \:.
\end{align}
Inserting the final condition for $\hat{p}$~\eqref{eq:BoundCdtGenCurrent} in the last term, we obtain
\begin{align}
    \dt{}{T}  \psi(\lambda; \nu,T)
    =& {} \:
    \frac{1}{2 \sqrt{T}} \lambda \int_0^\infty \dd x
        \partial_x \left(
        \xi \hat{q}(\xi \sqrt{T} + x, T;T) - \xi_0\hat{q}(\xi_0\sqrt{T} + x,0;T)
        \right)
    \nonumber
    \\
    &{}\:
    - \frac{1}{2 \sqrt{T}} \lambda 
     \rho_+ \left( \xi - \xi_0 \right)
    - \frac{\nu}{2 T^{3/2}} \int_{-\infty}^\infty \dd x \int_0^T \dd t \: \sigma(\hat{q}) \partial_x \hat{p} 
    \nonumber
    \\
    &{}\: - \int_{-\infty}^\infty \left[
    D(\hat{q}) \partial_x \hat{q} \partial_x \hat{p}
    - \frac{\nu}{\sqrt{T}} \sigma(\hat{q}) \partial_x \hat{p}
    - \frac{\sigma(\hat{q})}{2} (\partial_x \hat{p})^2
    \right]
    \Bigg|_{t=T}
    \dd x
    \:.
\end{align}
Similarly we have that
\begin{align}
    \dt{}{\nu}  \psi(\lambda; \nu,T)
    =& {} \:  \partial_\nu Q_T\left[\hat{q}; \xi, \xi_0 \right]
    - \partial_\nu S_T[\hat{q},\hat{p}]
    \nonumber 
    \\
    =& {} \:
    \frac{1}{\sqrt{T}}
    \int_{-\infty}^\infty \dd x \int_0^T \dd t \: \sigma(\hat{q}) \partial_x \hat{p} 
    \:.
\end{align}
Combining the two results, and going back to the original fields $(q,p)$ using~\eqref{eq:DefRescQP}, we obtain
\begin{multline}
    \dt{}{T}  \psi(\lambda; \nu,T)
    + \frac{1}{2T} \nu \dt{}{\nu}  \psi(\lambda; \nu,T)
    = -\frac{1}{2\sqrt{T}} \lambda
    \rho_+ \left( \xi - \xi_0 \right)
    +\frac{1}{2 \sqrt{T}} \lambda \int_0^\infty \dd x \:
        \partial_x \left(
        \xi q(\xi + x, 1) - \xi_0 q(\xi_0 + x,0)
        \right)
        \\
    - \frac{1}{\sqrt{T}} \int_{-\infty}^\infty \left[
    \frac{ 2 D(q)}{\sigma(q)} \partial_x q 
    - 2 \nu
    -  \partial_x p
    \right]
    \frac{\sigma(q)}{2} \partial_x p
    \Bigg|_{t=1}
    \dd x
    \:.
    \label{eq:RelGenCumul0}
\end{multline}
The integrals in the second line cannot be easily computed, since $q(x,1)$ is discontinuous at $\xi$ and $q(x,0)$ is discontinuous at $\xi_0$: the spatial derivative involves Dirac delta functions at the boundary of the integral, so their contributions is not well defined. To avoid this difficulty, we rewrite this term as
\begin{align}
    \lambda \int_0^\infty \dd x \:
        \partial_x \big(
        & \xi q(\xi + x, 1) - \xi_0q(\xi_0+ x,0)
        \big)
        \nonumber
        \\
    &= \lambda \int_{-\infty}^\infty \dd x \left[
    \Theta(x-\xi)
    \xi
    \partial_x  q( x, 1) -
    \Theta(x - \xi_0 )
    \xi_0\partial_x q(x,0)
    \right]
    \nonumber
    \\
    &= - \lambda \int_{-\infty}^\infty \dd x \left[
    \delta(x-\xi)
    \xi 
    q( x, 1) -
    \delta(x - \xi_0  )
    \xi_0 q(x,0)
    \right]
    \nonumber
    \\
    &{} \quad + \lambda (\xi - \xi_0) \rho_+
    \nonumber
    \\
    &= - \int_{-\infty}^\infty x \Big[ [q \partial_x p]_{t=0}^{t=1}
    + q \partial_x \mu(q) \big|_{t=0} - q(x,0) \partial_x \mu(\bar\rho(x))
    \Big] \dd x
    + \lambda (\xi - \xi_0) \rho_+
    \label{eq:RewritingIntegGenCurrent}
    \:,
\end{align}
where we have used the initial and final conditions on $p$~\eqref{eq:BoundCdtGenCurrent} in the last equality. The similarity with the terms obtained in~\eqref{eq:CumulNew2}, which was obtained by a different method, validates this rewriting. Since $\partial_x \bar\rho(x) \propto \delta(x)$, this term vanishes due to the $x$ in prefactor. Using the expression~\eqref{eq:RewritingIntegGenCurrent} into~\eqref{eq:RelGenCumul0}, together with the scaling of $\psi$~\eqref{eq:ScaledPsiGenCurrent} we obtain
\begin{equation}
    \label{eq:GenCumulQ1}
    \hat\psi
    + \nu \partial_\nu \hat\psi
    = 
    - \int_{-\infty}^\infty x q \partial_x p \Big|_{t=1} \dd x
    + \int_{-\infty}^\infty x q \left[ \partial_x p - \partial_x \mu(q) \right] \Big|_{t=0}
   \dd x
    - \int_{-\infty}^\infty \left[
    \partial_x \mu(q) 
    - 2 \nu
    -  \partial_x p
    \right]
    \sigma(q) \partial_x p
    \Big|_{t=1}
    \dd x
    \:.
\end{equation}
To analyse these terms, we again introduce the function $r = p - \mu(q)$, so that the MFT equations~\eqref{eq:MFTbulkDrive} become
\begin{align}
    \label{eq:MFTbulkRDriveGenQ}
    \partial_t r 
    &= D(q) \partial_x^2 r - \frac{\sigma'(q)}{2} (\partial_x r)^2 - \nu \sigma'(q) \partial_x r
    \:,
    &
    \partial_t p &= - D(q) \partial_x^2 p - \frac{\sigma'(q)}{2} (\partial_x p)^2 - \nu \sigma'(q) \partial_x p
    \:,
    \\
    \label{eq:MFTboundRDriveGenQ}
    r(x,0) &= \lambda \Theta \left(x - \xi_0 \right) - \mu(\bar\rho(x))
    \:,
    &
    p(x,1) &=  \lambda \Theta \left(x - \xi \right)
    \:.
\end{align}
This implies that $\partial_x (\mu(q) - p) = -\partial_x r$ is smooth for $t>0$ since it is solution of a diffusion equation, and $p(x,t)$ is smooth for $t<1$. Additionally, $\partial_x p(x,1)$ and $\partial_x r(x,0)$ are both sums or Dirac delta functions, so Eq.~\eqref{eq:GenCumulQ1} becomes
\begin{multline}
    \label{eq:PsiGenCumulQ2}
    \hat\psi
    + \nu \partial_\nu \hat\psi
    = 
    -  \xi \int_{\xi^{-}}^{\xi^{+}}
    \mu^{-1} \big( p(x,1) - r(\xi,1) \big) \partial_x p(x,1) \dd x
    +  \xi_0 \int_{\xi_0^{-}}^{\xi_0^{+}}
    \mu^{-1} \big( p(\xi_0 ,0) - r(x,1) \big) \partial_x r(x,0) \dd x
    \\
    - \big( \partial_x \mu(q) - 2 \nu \big) \Big|_{x=\xi}
    \int_{\xi^{-}}^{\xi^{+}} \sigma\left( \mu^{-1} \big( p(x,1) - r(\xi,1) \big) \right) \partial_x p(x,1)\dd x
    \Bigg]
    \:.
\end{multline}
The integral in the last term is given by~\eqref{eq:IntegT1}, while the other integrals can be expressed in terms of the pressure $P(\rho)$, defined by
\begin{equation}
    \label{eq:DefPressure}
    P'(\rho) = \rho \: \mu'(\rho) = \frac{2 \rho D(\rho)}{\sigma(\rho)}
    \:.
\end{equation}
Indeed,
\begin{equation}
    \partial_x P \left( \mu^{-1} \big( p(\xi_0 ,0) - r(x,1) \big) \right)
    =- \frac{P'\left( \mu^{-1} \big( p(\xi_0 ,0) - r(x,1) \big) \right)}
    {\mu'\left( \mu^{-1} \big( p(\xi_0 ,0) - r(x,1) \big) \right)} \partial_x r(x,1)
    =- \mu^{-1} \big( p(\xi_0 ,0) - r(x,1) \big) \partial_x r(x,1)
    \:,
\end{equation}
\begin{equation}
    \partial_x P \left( \mu^{-1}  \big( p(x,1) - r(\xi,1) \big) \right)
    = \frac{P'\left( \mu^{-1}  \big( p(x,1) - r(\xi,1) \big) \right)}
    {\mu'\left( \mu^{-1}  \big( p(x,1) - r(\xi,1) \big) \right)} \partial_x p(x,1)
    = \mu^{-1}  \big( p(x,1) - r(\xi,1) \big) \partial_x p(x,1)
    \:.
\end{equation}
Using these expressions into~\eqref{eq:PsiGenCumulQ2}, we finally obtain the expression of the shortcut for the cumulant generating function the generalised current~\eqref{eq:DefGenIntegCurrent0},
\begin{empheq}[box = \fbox]{multline}
    \label{eq:PsiGenCumulQFinal0}
    \hat\psi
    + \nu \partial_\nu \hat\psi
    =- 
      \xi\left[
    P \left( q \big( \xi^{+},1 \big) \right)
    -
    P \left( q \big( \xi^{-},1 \big) \right)
    \right]
    - \xi_0 \left[
    P \left( q \big( \xi_0^{+},0 \big) \right)
    - 
    P \left( q \big( \xi_0^{-},0 \big) \right)
    \right]
    \\
    - 2 \big( \partial_x \mu(q) - 2 \nu \big) \Big|_{x=\xi}
    \int_{q(\xi^{-},1)}^{q(\xi^{+},1)} D(r) \dd r 
    \:.
\end{empheq}
Remarkably, this expression involves only physical quantities: the pressure $P(\rho)$, the chemical potential $\mu(\rho)$ and the diffusion coefficient $D(\rho)$.

Note that an equation for $\hat\psi$ can be written in terms of the profile $q(x,1)$ at final time only, by noticing that
\begin{equation}
    \dt{}{\xi} \psi(\lambda;\nu,T) 
    = \lambda \partial_\xi Q_T[\hat{q}; \xi;\xi_0]
    = \lambda \sqrt{T} \left[ 
    \int_0^\infty \partial_x \hat{q}(\xi \sqrt{T} + x , T; T)
    - \rho_+
    \right]
    \:.
\end{equation}
The function $\hat{q}(x,T;T)$ is discontinuous at $x = \xi$, so the integral above contains a Dirac delta function on the boundary of the integration domain. Regularising it at in~\eqref{eq:RewritingIntegGenCurrent}, we finally get that
\begin{equation}
    \xi \dt{}{\xi} \psi(\lambda;\nu,T) 
    = - \xi \sqrt{T} \left[
    P \left( q \big( \xi^{+},1 \big) \right)
    -
    P \left( q \big( \xi^{-},1 \big) \right)
    \right]
    \:.
\end{equation}
Similarly
\begin{equation}
    \xi_0 \dt{}{\xi} \psi(\lambda;\nu,T) 
    = - \xi \sqrt{T} \left[
    P \left( q \big( \xi_0^{+},0 \big) \right)
    -
    P \left( q \big( \xi_0^{-},0 \big) \right)
    \right]
    \:,
\end{equation}
so that the cumulant generating function~\eqref{eq:PsiGenCumulQFinal0} can also be written as
\begin{equation}
    \label{eq:PsiGenCumulQFinal}
    \boxed{
    \hat\psi + \nu \partial_\nu \hat\psi
    - \xi \partial_\xi \hat\psi - \xi_0 \partial_{\xi_0} \hat\psi
    = - 2 \big( \partial_x \mu(q) - 2 \nu \big) \Big|_{x=\xi}
    \int_{q(\xi^{-},1)}^{q(\xi^{+},1)} D(r) \dd r 
    \:.
    }
\end{equation}
This expression shows that the determination of the profile $q(x,1)$ at the final time is sufficient to determine the cumulant generating function of the integrated current. We will illustrate this in Section~\ref{sec:Applications} below.

\subsection{Case of a quenched initial condition}

Importantly, we have considered up to now the case of an annealed cumulant generating function, for instance defined by~\eqref{eq:CumulGenCurrent} for the joint study of integrated currents. We can also consider a quenched situation, where
\begin{equation}
    \label{eq:CumulCurrentQuenched}
    \hat\psi_{\mathrm{qu.}}(\lambda; \nu) 
    \equiv \lim_{T \to \infty} \frac{1}{\sqrt{T}}
   \moy{ \ln \moy{
    \exp \left[
     \lambda  Q_T[x]
    \right]
    }_{\mathrm{e}}}_{\mathrm{i}}
    \:,
\end{equation}
In this case, one has to solve the same MFT equations~\eqref{eq:MFTbulkDrive}, but with the modified boundary conditions
\begin{equation}
    \label{eq:BoundCdtGenCurrentQuenched}
    p(x;1) = \lambda \Theta \left( x - \xi \sqrt{T} \right)
    \:,
    \quad
    q(x,0) = \bar\rho(x)
    \:,
\end{equation}
with $\bar\rho(x)$ the step of density defined in~\eqref{eq:StepInit}.
We can perform the exact same derivation as in the previous section up to~\eqref{eq:RelGenCumul0}, which gives after inserting $q(x,0) = \bar\rho(x)$,
\begin{multline}
    \dt{}{T}  \psi_{\mathrm{qu.}}(\lambda; \nu,T)
    + \frac{1}{2T} \nu \dt{}{\nu}  \psi_{\mathrm{qu.}}(\lambda; \nu,T)
    = -\frac{1}{2\sqrt{T}}\lambda 
    \rho_+ \left( \xi - \xi_0 \right)
    +\frac{1}{2 \sqrt{T}} \lambda_i \int_0^\infty \dd x \:
        \partial_x \left(
        \xi q(\xi + x, 1) - \xi_0\bar\rho(\xi_0 + x)
        \right)
        \\
    - \frac{1}{\sqrt{T}} \int_{-\infty}^\infty \left[
    \frac{ 2 D(q)}{\sigma(q)} \partial_x q 
    - 2 \nu
    -  \partial_x p
    \right]
    \frac{\sigma(q)}{2} \partial_x p
    \Bigg|_{t=1}
    \dd x
    \:,
    \label{eq:RelJointCumulQuenched0}
\end{multline}
where $\psi_{\mathrm{qu.}}(\lambda; \nu,T) = \sqrt{T} \hat\psi_{\mathrm{qu.}}(\lambda; \nu)$.
The second term in~\eqref{eq:RelJointCumulQuenched0} is again more conveniently rewritten as in~\eqref{eq:RewritingIntegGenCurrent}, which here gives
\begin{equation}
   \lambda \int_0^\infty \dd x \:
        \partial_x \big(
         \xi q(\xi + x, 1) - \xi_0 q(\xi_0 + x,0)
        \big)
    = - \int_{-\infty}^\infty x  q \partial_x p \Big|_{t=1} \dd x
    +  \lambda \left[ (\xi - \xi_0) \rho_+
    + \xi_0 \bar\rho(\xi_0)
    \right]
    \label{eq:RewritingIntegGenCurrentQuenched}
    \:.
\end{equation}
Therefore,
\begin{equation}
    \hat\psi_{\mathrm{qu.}}
    + \nu \partial_\nu \hat\psi_{\mathrm{qu.}}
    = \lambda \xi_0 \bar\rho(\xi_0)
    -  \int_{-\infty}^\infty x  q \partial_x p \Big|_{t=1} \dd x
    - \int_{-\infty}^\infty \left[
    \frac{ 2 D(q)}{\sigma(q)} \partial_x q 
    - 2 \nu
    -  \partial_x p
    \right]
    \sigma(q) \partial_x p
    \Bigg|_{t=1}
    \dd x
    \:.
\end{equation}
Repeating the analysis of the integrals at $t=1$ from the previous section, we obtain the joint cumulant generating function in the quenched case
\begin{equation}
    \label{eq:PsiGenCumulQFinalQuenched}
    \boxed{
    \hat\psi_{\mathrm{qu.}}
    + \nu \partial_\nu \hat\psi_{\mathrm{qu.}}
    =- 
      \xi \left[
    P \left( q \big( \xi^{+},1 \big) \right)
    -
    P \left( q \big( \xi^{-},1 \big) \right)
    \right]
    + \lambda \xi_0 \bar\rho(\xi_0)
    - 2 \big( \partial_x \mu(q) - 2 \nu \big) \Big|_{x=\xi}
    \int_{q(\xi^{-},1)}^{q(\xi^{+},1)} D(r) \dd r 
    }
    \:.
\end{equation}
This expression in the quenched case is almost identical to the one in the annealed case~\eqref{eq:PsiGenCumulQFinal}: it involves the same contributions of the discontinuities of the profile at the final time $t=1$. The discontinuities of the pressure at the initial time involved in~\eqref{eq:PsiGenCumulQFinal} have disappeared in~\eqref{eq:PsiGenCumulQFinalQuenched} because the initial profile is now fixed and thus cannot be discontinuous at arbitrary positions. However, an additional term $\lambda \xi_0 \bar\rho(\xi_0)$ involving the initial condition has appeared. Its origin can be identified easily: since the initial profile is fixed, the generalised current~\eqref{eq:DefGenIntegCurrentMacroFinal} can be written as
\begin{equation}
     Q[\rho; \xi, \xi_0] =
     Q[\rho; \xi, 0] + \xi_0 \bar\rho(\xi_0)
\end{equation}
because the number of particles between $0$ and $\xi_0$ is deterministic. The term $\lambda \xi_0 \bar\rho(\xi_0)$ in~\eqref{eq:PsiGenCumulQFinalQuenched} simply comes from this deterministic contribution to the generalised currents in the quenched case.

\subsection{Local time}

Up to now, we have considered observables that involve integrated currents. We consider now another type of observable, the local time at a given position $y$,
\begin{equation}
    \label{eq:localtime0}
    R_T(y) = \int_0^T\rho_0(y,t) \dd t
    \:,
\end{equation}
which corresponds to the total time the position $y$ has been occupied by any particle. This quantity has recently been studied within the MFT formalism~\cite{Meerson:2024}. It was shown that the distribution of $R_T(y)$ takes the large deviation form for $T \to \infty$,
\begin{equation}
    P(R_T(y)) \simeq \e^{- \sqrt{T} \phi( R_T(y)/T)}
    \:.
\end{equation}
Performing a Laplace transform, this implies that the cumulant generating function takes the form
\begin{equation}
    \label{eq:CumulLocalTime}
    \ln \moy{ \moy{\e^{\chi R_T(y)} }_{\mathrm{e}} }_{\mathrm{i}}
    \underset{T \to \infty}{\simeq}
    \sqrt{T} \: \hat{\psi}\left( \frac{\chi}{\sqrt{T}} \right)
    \:,
\end{equation}
where we have now used the notation $\chi$ for the generating function of the cumulants to distinguish it with the case of the currents considered above. 
In terms of the macroscopic fields~\eqref{eq:DefMacroFields} with rescaling parameter $\Lambda = \sqrt{T}$, the local time can be expressed as
\begin{equation}
    \label{eq:localtime}
    R_T(y) = T \:
    \mathcal{R}_\zeta[\rho] 
    \:,
    \quad
    \mathcal{R}_\zeta[\rho] \equiv
    \int_0^1 \rho(\zeta,t) \dd t
    \:,
\end{equation}
where we have introduced $\zeta = y/\sqrt{T}$. The scaled cumulant generating function $\hat\psi$~\eqref{eq:CumulLocalTime} can be computed within the MFT formalism as
\begin{equation}
    \hat\psi(\chi) = 
    \chi \mathcal{R}_\zeta[q] - S[q,p] - F[q(x,0)]
    \:,
\end{equation}
with the MFT action given by~\eqref{eq:MFTactionDrive} and $F$ is expressed as~\eqref{eq:DistInit}. The fields $(q,p)$ minimise $\chi \mathcal{R}_\zeta - S - F$ and obey the modified MFT equations~\cite{Meerson:2024}
\begin{equation}
    \label{eq:MFTLocalTime}
  \partial_t q 
  =
  \partial_{x} \left[D(q)\partial_{x}q- \sigma(q)\partial_{x}p - \nu \sigma(q)
  \right]
  \:,
  \quad
    \partial_{t}p =
    -D(q)\partial_{x}^{2}p
    -\frac{1}{2}\sigma'(q)(\partial_{x}p)^{2}
    - \nu \sigma'(q) \partial_x p
    -\chi\delta(x-\zeta)
    \:,
\end{equation}
with the initial and final conditions
\begin{equation}
\label{eq:MFTinitfinLocalTime}
    p(x,1) = 0
    \:,
    \quad
    p(x,0) = \int_{\bar\rho(x)}^{q(x,0)} \frac{2 D(r)}{\sigma(r)} \dd r
    \:.
\end{equation}
As for the generalised current, it is useful to introduce the rescaled fields~\eqref{eq:DefRescQP}, from which we can express the cumulant generating function as
\begin{equation}
    \label{eq:CumulLocalTimeNew}
    \ln \moy{ \moy{\e^{\chi R_T(\zeta \sqrt{T}) / \sqrt{T}} }_{\mathrm{e}} }_{\mathrm{i}}
    \underset{T \to \infty}{\simeq}
    \sqrt{T} \: \hat{\psi}(\chi) \equiv
    \psi(\chi;T)
    \:,
\end{equation}
where
\begin{equation}
    \label{eq:psiLocalTimeT}
    \psi(\chi;T) = \frac{\chi}{\sqrt{T}} R_T[\hat{q}] - S_T[\hat{q},\hat{p}] - F[\hat{q}(x,0)]
    \:,
\end{equation}
with $S_T$ given by~\eqref{eq:ActionT} and
\begin{equation}
    R_T[\hat{q}] \equiv \int_0^T \hat{q}( \zeta \sqrt{T}, t; T) \dd t
    \:.
\end{equation}
Using that $(\hat{q},\hat{p})$ minimise~\eqref{eq:psiLocalTimeT}, we can compute
\begin{align}
    \dt{}{T} \psi(\chi; T) 
    &= {}\: \partial_T \left( \frac{\chi}{\sqrt{T}} R_T[\hat{q}] - S_T[\hat{q},\hat{p}] \right)
    \nonumber
    \\
    &= - \frac{1}{2 T^{3/2}} R_T[\hat{q}]
    + \frac{\chi}{\sqrt{T}} \hat{q}(\zeta \sqrt{T}, T; T)
    + \frac{\chi \zeta}{2 T} \int_0^T \partial_x \hat{q}(\zeta \sqrt{T},t;T) \dd t
    - \frac{\nu}{2 T^{3/2}} \int_{-\infty}^\infty \dd x \int_0^T \dd t \: \sigma(\hat{q}) \partial_x \hat{p}
    \nonumber
    \\
    & {}\: - \int_{-\infty}^\infty \dd x \left[
    \hat{p} \partial_t \hat{q} + D(\hat{q}) \partial_x \hat{q} \partial_x \hat{p}
    - \frac{\nu}{\sqrt{T}} \sigma(\hat{q}) \partial_x \hat{p}
    - \frac{\sigma(\hat{q})}{2} (\partial_x \hat{p})^2
    \right] \Bigg|_{t=T}
    \:.
\end{align}
Reintroducing the original MFT solution $(q,p)$~\eqref{eq:DefRescQP} and using the final condition of $p$~\eqref{eq:MFTinitfinLocalTime}, this becomes
\begin{equation}
    \dt{}{T} \psi(\chi; T) 
    =  -\frac{\chi}{2 \sqrt{T}} \mathcal{R}_\zeta[q]
    + \frac{\chi}{\sqrt{T}} q(\zeta, 1)
    + \frac{\chi \zeta}{2 \sqrt{T}} \int_0^1 \partial_x q(\zeta,t) \dd t
    - \frac{\nu}{2 \sqrt{T}} \int_{-\infty}^\infty \dd x \int_0^1 \dd t \: \sigma(q) \partial_x p
    \:.
\end{equation}
Similarly, we have that
\begin{equation}
    \dt{}{\chi} \psi(\chi; T) 
    =  \frac{1}{\sqrt{T}} R_T[\hat{q}]
    = \sqrt{T} \mathcal{R}_\zeta[q]
    \:,
    \quad
    \dt{}{\nu} \psi(\chi; T) =\frac{1}{\sqrt{T}} \int_{-\infty}^\infty \dd x \int_0^T \dd t \: \sigma(\hat{q}) \partial_x \hat{p}
    \:,
    \quad
    \dt{}{\zeta} \psi(\chi; T) =
    \sqrt{T} \chi \int_{0}^{1} \partial_x q(\zeta,t) \dd t
    \:.
\end{equation}
Combining these expressions and using the scaling form~\eqref{eq:CumulLocalTimeNew}, we deduce that
\begin{equation}
    \label{eq:ShortcutLocalTime}
    \boxed{
    \hat\psi + \nu \partial_\nu \hat\psi + \chi \partial_\chi \hat\psi
    - \zeta \partial_\zeta \hat\psi
    = 2 \chi q(\zeta,1) 
    \:.
    }
\end{equation}
This relation shows that the cumulant generating function of the local time is again fully determined by the value of the MFT profile at final time, and only at $x=\zeta$.
Finally note that the derivation above can be straightforwardly extended to the quenched case, so that the relation~\eqref{eq:ShortcutLocalTime} applies both to quenched and annealed initial conditions.

\subsection{Joint distribution of several observables}

We have up to now considered different observables, both integrated currents and local times, but always studied a single observable at a time. These different quantities are however correlated, and it is natural to study their joint statistical properties. This is the object of this section.

We consider the joint distribution of $N$ integrated currents of the form~\eqref{eq:DefGenIntegCurrentMacroFinal} and of $M$ local times~\eqref{eq:localtime}. All the statistical properties are encoded in the joint cumulant generating function
\begin{equation}
    \label{eq:JointCumul}
    \hat\psi(\{ \lambda_i \}, \{ \chi_j \}; \nu) 
    \equiv \lim_{T \to \infty} \frac{1}{\sqrt{T}}
    \ln \moy{\moy{
    \exp \left[
    \sqrt{T} \sum_{i=1}^N \lambda_i  Q \left[\rho;\xi^{(i)}, \xi_0^{(i)} \right]
    + \sqrt{T} \sum_{j=1}^M \chi_j \mathcal{R}_{\zeta^{(j)}}[\rho]
    \right]
    }_{\mathrm{e}}}_{\mathrm{i}}
    \:,
\end{equation}
where $\xi^{(i)}$ and $\xi_0^{(i)}$ are the final and initial positions at which the $i^{\mathrm{th}}$ current is measured (up to a rescaling by $\sqrt{T}$) and $\zeta^{(j)}$ is the position where the $j^{\mathrm{th}}$ local time is measured.
In this case, the MFT equations become
\begin{equation}
    \label{eq:MFTjoint}
  \partial_t q 
  =
  \partial_{x} \left[D(q)\partial_{x}q- \sigma(q)\partial_{x}p - \nu \sigma(q)
  \right]
  \:,
  \quad
    \partial_{t}p =
    -D(q)\partial_{x}^{2}p
    -\frac{1}{2}\sigma'(q)(\partial_{x}p)^{2}
    - \nu \sigma'(q) \partial_x p
    - \sum_{j=1}^M \chi_j \delta \left(x-\zeta^{(j)} \right)
    \:,
\end{equation}
with the initial and final conditions
\begin{equation}
\label{eq:MFTinitfinjoin}
    p(x,1) = \sum_{i=1}^N \lambda_i \Theta \left( x - \xi^{(i)} \right)
    \:,
    \quad
    p(x,0) =
    \sum_{i=1}^N \lambda_i \textbf{$\Theta$} \left( x - \xi_0^{(i)} \right)
    +
    \int_{\bar\rho(x)}^{q(x,0)} \frac{2 D(r)}{\sigma(r)} \dd r
    \:.
\end{equation}

The derivations of the previous Sections can be directly transposed here to yield
\begin{empheq}[box = \fbox]{multline}
    \label{eq:PsiJointCumulFinal}
    \hat\psi
    + \nu \partial_\nu \hat\psi
    + \sum_{j=1}^M \left( \chi_j \partial_{\chi_j} \hat\psi
    - \zeta^{(j)} \partial_{\zeta^{(j)}} \hat\psi
    \right)
    =
    \sum_{j=1}^M 
    2 \chi_j q(\zeta^{(j)},1)
    \\
    - \sum_{i=1}^N \Bigg[ 
      \xi^{(i)} \left[
    P \left( q \big( \xi^{(i)+},1 \big) \right)
    -
    P \left( q \big( \xi^{(i)-},1 \big) \right)
    \right]
    + \xi_0^{(i)} \left[
    P \left( q \big( \xi_0^{(i)+},0 \big) \right)
    - 
    P \left( q \big( \xi_0^{(i)-},0 \big) \right)
    \right]
    \\
    + 2 \big( \partial_x \mu(q) - 2 \nu \big) \Big|_{x=\xi^{(i)}}
    \int_{q(\xi^{(i)-},1)}^{q(\xi^{(i)+},1)} D(r) \dd r 
    \Bigg]
    \:.
\end{empheq}
Remarkably, this joint cumulant generating function is expressed as a sum over the contributions of the solution of the MFT equations for each individual observable, as if they were independent. This is of course not the case, because they are coupled through the MFT profile $q(x,t)$ which depends on all the parameters $\lambda_i$ and $\zeta_j$.

We stress that the expression~\eqref{eq:PsiJointCumulFinal} expresses the joint scaled cumulant generating function~\eqref{eq:JointCumul} in terms of the solution of the MFT equations at the final time $q(x,1)$ and initial time $q(x,0)$ only. This expression also involves only physical quantities: the pressure $P(\rho)$, the chemical potential $\mu(\rho)$ and the diffusion coefficient $D(\rho)$.

\subsection{Finite systems}
\label{sec:finite}

Up to now we have only considered infinite systems, but the previous analysis can be extended to other geometries such as a finite one-dimensional system of large length \( L \gg 1 \), connected at its boundaries to two particle reservoirs with densities \( \rho_L \) on the left and \( \rho_R \) on the right.
It is the paradigmatic setting of out-of-equilibrium systems~\cite{Derrida:2025a}, that has been the focus of many studies~\cite{Bodineau:2004,Derrida:2004,Bodineau:2008,Akkermans:2013,Bodineau:2025,Saha:2024,Saha:2025}.
In such finite systems, both the current and the density reach a stationary state at long times.
In this case, the natural rescaling parameter to go from the microscopic to the macroscopic scale in~\eqref{eq:DefMacroFields} is $\Lambda = L$. 
We consider the time-integrated current through an arbitrary point $x = \xi L$ with \( \xi \in [0, 1] \), defined by
\begin{equation}
    Q_T(x) = \int_0^T j_0(x, t)\, \dd t
    \:,
    \label{eq:currentFinite0}
\end{equation}
and the local time at a given position $y =  \zeta L$, $\zeta \in [0,1]$,
\begin{equation}
    \label{eq:LocalFinite0}
    R_T(y) = \int_0^T \rho_0(y,t) \dd t
    \:.
\end{equation}
Performing the scaling~\eqref{eq:DefMacroFields} to the macroscopic case, we rewrite these observables as
\begin{equation}
    Q_T(x) = L \int_0^{T/L^2} j(\xi,t) \dd t
    \:,
    \quad
    R_T(y) = L^2 \int_0^{T/L^2} \rho(\zeta,t) \dd t
    \:.
\end{equation}
We consider the joint cumulant generating function of $N$ generalised currents~\eqref{eq:currentFinite0} and $M$ local times~\eqref{eq:LocalFinite0}, defined as
\begin{equation}
    \label{eq:DefJointFinite}
    \lim_{L \to \infty} \frac{1}{L} \ln \moy{ \moy{
    \exp \left[
        \sum_{i=1}^N \lambda_i Q_{L^2 T}(x_i)
        + \sum_{j=1}^M \chi_j \frac{R_{L^2 T}(y_j)}{L}
    \right]
    }_{\mathrm{e}} }_{\mathrm{i}}
    \equiv \psi \left(
    \{ \lambda_i \}, \{ \chi_j \}, T
    \right)
    \:,
\end{equation}
where we have rescaled the local times by $L$ to ensure the correct scaling of the cumulants, and scaled the measurement time with $L^2$. We have denoted $x_i = \xi^{(i)} L$ the position where the $i^{\mathrm{th}}$ current is measured, and $y_i = \zeta^{(j)} L$ the position where the $j^{\mathrm{th}}$ local time is measured. The MFT formalism still applies in this case~\cite{Bodineau:2004} and yields
\begin{equation}
    \psi \left(
    \{ \lambda_i \}, \{ \chi_j \}, T
    \right)
    = \sum_{i=1}^N \lambda_i Q_{T}[q,p; \xi^{(i)}] + 
    \sum_{j=1}^M \chi \mathcal{R}_{T}[q; \zeta^{(i)}]
    - S_T[q,p] - F[q(x,0)]
    \:,
\end{equation}
where
\begin{equation}
    \label{eq:DefQTmacroFinite}
    Q_T[q,p;\xi] = \int_0^T j(\xi,t) \dd t 
    \:,
    \quad \text{where} \quad
    j = - D(q) \partial_x q - \sigma(q) \partial_x p
    \:,
\end{equation}
\begin{equation}
    \mathcal{R}_T[q;\zeta]
    = \int_0^T q(\zeta,t) \dd t
    \:.
\end{equation}
The distribution of the initial condition $F$ will be irrelevant here, because the system reaches the same steady state at long time for any initial condition, so both quenched and annealed case can be treated in the same way and we do not need to specify $F$. 
The MFT action $S_T[q,p]$ is now given by
\begin{equation}
    \label{eq:MFTactionDriveFinite}
    S_T[q,p] = \int_{0}^1 \dd x \int_0^T \dd t \left[
    p \partial_t q + D(q) \partial_x q \partial_xp
    - \nu \sigma(q) \partial_x p
    - \frac{\sigma(q)}{2} (\partial_x p)^2
    \right]
    + \int_{0}^T (j(1,t) p(1,t) - j(0,t) p(0,t)) \dd t
    \:.
\end{equation}
Note that here, $\nu$ does not depend on $T$ because the drive has been rescaled by the large size $L$ of the system, and not by $\sqrt{T}$.
The evolution equations of $(q,p)$ are again obtained from minimising the action. The only difference here is that we cannot rewrite $Q_T$ in terms of integrals of the density only, so we must keep the current $j$, related to $(q,p)$ by~\eqref{eq:DefQTmacroFinite}. This yield the same MFT equations
\begin{equation}
    \label{eq:MFTjointFinite}
  \partial_t q 
  =
  \partial_{x} \left[D(q)\partial_{x}q- \sigma(q)\partial_{x}p - \nu \sigma(q)
  \right]
  \:,
  \quad
    \partial_{t}p =
    -D(q)\partial_{x}^{2}p
    -\frac{1}{2}\sigma'(q)(\partial_{x}p)^{2}
    - \nu \sigma'(q) \partial_x p
    - \sum_{j=1}^M \chi_j \delta \left(x-\zeta^{(j)} \right)
    \:,
\end{equation}
but restricted to $x \in [0,1]$ (and $x \neq \xi^{(i)}$) and $t \in [0,T]$, and completed by the boundary conditions at the reservoirs~\cite{Saha:2023,Saha:2024}
and in space
\begin{equation}
    q(0, t) = \rho_L
    \:, 
    \quad 
    q(1, t) = \rho_R
    \:, 
    \quad p(0, t) = 0
    \:, 
    \quad 
    p(1, t) = 0
    \:.
    \label{eq:BC-space}
\end{equation}
The final condition in time now becomes
\begin{equation}
    p(x,1) = 0
    \:.
\end{equation}
As we will see, the initial condition will play no role, so we do not specify it. The main difference is that now, $p$ is no longer continuous, but instead satisfies
\begin{equation}
    p(\xi^{(i)+},t) - p(\xi^{(i)-},t) + \lambda_i = 0
    \:.
\end{equation}
We can reduce this apparently different problem to the usual formulation considered above by redefining $p$ as
\begin{equation}
    p(x,t) \longrightarrow p(x,t) - \sum_{i=1}^N \lambda_i \Theta(x - \xi^{(i)})
    \:,
\end{equation}
so that the new field $p$ is continuous, and obeys the same MFT equations~\eqref{eq:MFTjointFinite}, with now
\begin{equation}
    \label{eq:FinalCondFinite}
    p(x,1) = \sum_{i=1}^N \lambda_i \Theta(x - \xi^{(i)})
    \:,
    \quad
    p(1,t) = \sum_{i=1}^N \lambda_i
    \:.
\end{equation}
With these boundary conditions, the MFT action~\eqref{eq:MFTactionDriveFinite} now reads
\begin{equation}
    \label{eq:MFTactionDriveFiniteNew}
    S_T[q,p] = \int_{0}^1 \dd x \int_0^T \dd t \left[
    p \partial_t q + D(q) \partial_x q \partial_xp
    - \nu \sigma(q) \partial_x p
    - \frac{\sigma(q)}{2} (\partial_x p)^2
    \right]
    + \sum_{i=1}^N \lambda_i \int_{0}^T j(1,t)  \dd t
    \:.
\end{equation}
As in the infinite case, we compute the derivative of $\psi$ with respect to $T$,
\begin{align}
    \dt{}{T} \psi \left(
    \{ \lambda_i \}, \{ \chi_j \}, T
    \right)
    =& {}\: \partial_T \left(
     \sum_{i=1}^N \lambda_i Q_{T}[q,p; \xi^{(i)}] + 
    \sum_{j=1}^M \chi_i \mathcal{R}_{T}[q; \zeta^{(i)}]
    - S_T[q,p]
    \right)
    \nonumber
    \\
    =& {}\:
    \sum_{i=1}^N \lambda_i \left( j(\xi^{(i)},T) - j(1,T) \right)
    + \sum_{j=1}^M \chi_j q(\zeta^{(j)},T)
    \nonumber
    \\
    &{}\:
    - \int_{0}^1 \dd x \left[
    p \partial_t q + D(q) \partial_x q \partial_xp
    - \nu \sigma(q) \partial_x p
    - \frac{\sigma(q)}{2} (\partial_x p)^2
    \right] \Bigg|_{t=T}
    \:.
\end{align}
Inserting the final condition of $p$~\eqref{eq:FinalCondFinite} and combining with $\partial_t q = - \partial_x j$ and the boundary conditions~\eqref{eq:BC-space}, we obtain
\begin{equation}
    \dt{}{T} \psi \left(
    \{ \lambda_i \}, \{ \chi_j \}, T
    \right)
    = \sum_{j=1}^M \chi_j q(\zeta^{(j)},T)
    - \int_{0}^1 \dd x \left[
    \frac{2 D(q)}{\sigma(q)} \partial_x q
    - 2 \nu
    - \partial_x p
    \right] \frac{\sigma(q)}{2} \partial_x p \Bigg|_{t=T}
    \:.
\end{equation}
Performing the same analysis as in the previous Sections near the discontinuities of $p$ and $q$ at $x = \xi^{(i)}$, we obtain,
\begin{equation}
    \label{eq:ShortcutFiniteJoint}
    \boxed{
    \dt{}{T} \psi \left(
    \{ \lambda_i \}, \{ \chi_j \}, T
    \right)
    = \sum_{j=1}^M \chi_j q(\zeta^{(j)},T)
    - \sum_{i=1}^N \bigg( \partial_x \mu(q(x,T)) \big|_{x=\xi^{(i)}} - 2 \nu \bigg)
    \int_{q(\xi^{(i)-},T)}^{q(\xi^{(i)+},T)}
    D(r) \dd r
    \:.
    }
\end{equation}
In the long time limit, the joint cumulant generating function has the scaling behaviour
\begin{equation}
    \psi \left(
    \{ \lambda_i \}, \{ \chi_j \}, T
    \right)
    \underset{T \to \infty}{\simeq}
    T \: \hat\psi\left(
    \{ \lambda_i \}, \{ \chi_j \}
    \right)
    \:,
\end{equation}
which in~\eqref{eq:ShortcutFiniteJoint} yields
\begin{equation}
    \label{eq:ShortcutFiniteJointScaled}
    \hat\psi \left(
    \{ \lambda_i \}, \{ \chi_j \}
    \right)
    = \sum_{j=1}^M \chi_j q(\zeta^{(j)},T)
    - \sum_{i=1}^N \bigg( \partial_x \mu(q(x,T)) \big|_{x=\xi^{(i)}} - 2 \nu \bigg)
    \int_{q(\xi^{(i)-},T)}^{q(\xi^{(i)+},T)}
    D(r) \dd r
    \:.
\end{equation}
This formula is very similar to the one obtained in the infinite case~\eqref{eq:PsiJointCumulFinal}, except that compared to~\eqref{eq:PsiJointCumulFinal} some terms have vanished and the remaining terms differ by a factor $2$.

\section{Summary: a general formula for the shortcut}
\label{sec:UniversalShortcut}

We have obtained various formulas, depending on the observables under consideration and on the geometry of the system (infinite or finite), see for instance Eq.~\eqref{eq:PsiJointCumulFinal} for the infinite case and Eq.~\eqref{eq:ShortcutFiniteJointScaled} for the finite case. We now show that these two formulas come from the \textit{same} general formula that applies in all the cases considered here.

For this, let us go back to the original definition~\eqref{eq:JointCumul} of the joint cumulant generating function in the infinite case, but let us write it in terms of the microscopic parameters only
\begin{equation}
    \label{eq:JointCumulMicro}
    \psi_0 \left(\{ \lambda_i \}, \{ \chi_j \};
    \{ x_i \}, \{ y_j \};
    f ; T
    \right)
    \equiv 
    \ln \moy{\moy{
    \exp \left[
    \sum_{i=1}^N \lambda_i  Q_T[x_i]
    + \sum_{j=1}^M \chi_j R_T(y_j)
    \right]
    }_{\mathrm{e}}}_{\mathrm{i}}
    \:,
\end{equation}
where we have introduced the force $f$ driving the particles, and the positions $\{ x_i(t) \}$ and $\{ y_j(t) \}$ where the currents and local times are measured, respectively. To be consistent with the macroscopic scalings~\eqref{eq:DefMacroFields} and to get the correct scaling of the observables, we have previously studied the scaled joint cumulant generating function~\eqref{eq:JointCumul}, which is obtained from~\eqref{eq:JointCumulMicro} by
\begin{equation}
    \label{eq:RelMicroMacroJointCumul}
    \psi_0 \left(\{ \lambda_i \}, \left\lbrace \frac{\chi_j}{\sqrt{T}} \right\rbrace;
    \{ \bar{x}_{i;T} \}, \{ \bar{y}_{j;T} \};
    f = \frac{\nu}{\sqrt{T}} ; T
    \right)
    \underset{T \to \infty}{\simeq}
    \sqrt{T} \: 
    \hat\psi\left(\{ \lambda_i \}, \{ \chi_j \} \right)
    \:,
\end{equation}
where the positions are taken as
\begin{equation}
    \bar{x}_{i;T}(t) =
    \sqrt{T} \left( \xi_0^{(i)} + (\xi^{(i)}-\xi_0^{(i)}) \sqrt{t} \right)
    \:,
    \quad
    \bar{y}_{i;T} = \zeta^{(j)} \sqrt{T}
\end{equation}
to ensure that $\bar{x}_{i;T}(0) = \xi_0^{(i)} \sqrt{T}$ and $\bar{x}_{i;T}(T) = \xi^{(i)} \sqrt{T}$. Taking the time derivative~\eqref{eq:RelMicroMacroJointCumul}, we obtain that
\begin{equation}
    \dt{}{T} \psi_0 
    = \partial_T \psi_0
    - \frac{1}{2 \sqrt{T}} \nu \partial_\nu \hat\psi
    - \frac{1}{2\sqrt{T}} \sum_{j=1}^M \chi_j \partial_{\chi_j} \hat\psi
    + \frac{1}{2 \sqrt{T}}\sum_{j=1}^M \zeta^{(j)} \partial_{\zeta^{(j)}} \hat\psi
    + \frac{1}{2 \sqrt{T}} \sum_{i=1}^N \left( 
    \xi^{(i)} \partial_{\xi^{(i)}} \hat\psi
    + \xi_0^{(i)} \partial_{\xi_0^{(i)}} \hat\psi
    \right)
    \:.
\end{equation}
Therefore, we have
\begin{equation}
    2 \sqrt{T} \: \partial_T \psi_0
    = \hat\psi
    + \nu \partial_\nu \hat\psi
    + \sum_{j=1}^M \left( \chi_j \partial_{\chi_j} \hat\psi
    -  \zeta^{(j)} \partial_{\zeta^{(j)}} \hat\psi \right)
    - \sum_{i=1}^N \left( 
    \xi^{(i)} \partial_{\xi^{(i)}} \hat\psi
    + \xi_0^{(i)} \partial_{\xi_0^{(i)}} \hat\psi
    \right)
    \:.
\end{equation}
Using the definition of the generalised current~\eqref{eq:DefGenIntegCurrentMacroFinal}, we can compute
\begin{align}
    \dt{}{\xi^{(i)}} \hat\psi 
    &= {}\: \partial_{\xi^{(i)}} \left( 
    \lambda_i \int_0^\infty \left[ q(\xi^{(i)} + x, 1) - q(\xi_0^{(i)},0) \right] \dd x
    - \rho_+ \xi
    \right)
    \nonumber
    \\
    &= {}\: \lambda_i \int_0^\infty \partial_x q(\xi^{(i)}+x, 1) \dd x - \lambda_i \rho_+
    \:.
\end{align}
The function $q(x,1)$ is discontinuous at $x = \xi^{(i)}$, so the integral above contains a Dirac delta function on the boundary of the integration domain. Regularising it at in~\eqref{eq:RewritingIntegGenCurrent}, we finally get that
\begin{equation}
    \xi^{(i)} \partial_{\xi^{(i)}}\hat\psi 
    = - \xi^{(i)}  \left[
    P \left( q \big( \xi^{(i)+},1 \big) \right)
    -
    P \left( q \big( \xi^{(i)-},1 \big) \right)
    \right]
    \:,
\end{equation}
and similarly for $\xi_0^{(i)}$. Therefore, the joint scaled cumulant generating function~\eqref{eq:PsiJointCumulFinal} can also be written as
\begin{multline}
   2 \sqrt{T} \: \partial_T \psi_0 \left(\{ \lambda_i \}, \left\lbrace \frac{\chi_j}{\sqrt{T}} \right\rbrace;
    \{ \bar{x}_{i;T} \}, \{ \bar{y}_{j;T} \};
    f = \frac{\nu}{\sqrt{T}} ; T
    \right)
   = 
   2 \sum_{j=1}^M 
    \chi_j q(\zeta^{(j)},1)
    \\
    - 2\sum_{i=1}^N 
    \big( \partial_x \mu(q(x,1)) - 2 \nu \big) \Big|_{x=\xi^{(i)}}
    \int_{q(\xi^{(i)-},1)}^{q(\xi^{(i)+},1)} D(r) \dd r 
    \:.
\end{multline}
If we write this expression in terms of the microscopic parameters only, and in terms of the rescaled MFT field $\hat{q}$~\eqref{eq:DefRescQP} which depends explicitly on $T$, this equation becomes
\begin{empheq}[box = \fbox]{multline}
    \label{eq:UniversalShortcut}
    \partial_T \ln \moy{\moy{
    \exp \left[
    \sum_{i=1}^N \lambda_i  Q_T[x_i]
    + \sum_{j=1}^M \chi_j R_T(y_j)
    \right]
    }_{\mathrm{e}}}_{\mathrm{i}}
    \underset{T \to \infty}{\simeq}
    \sum_{j=1}^M 
    \chi_j \hat{q}(y_j,T)
    \\
    - \sum_{i=1}^N 
    \big( \partial_x \mu(\hat{q}(x,T)) - 2 \nu \big) \Big|_{x=x_i(T)}
    \int_{\hat{q}(x_i(T)^-,T)}^{\hat{q}(x_i(T)^+,T)} D(r) \dd r 
    \:.
\end{empheq}
Remarkably, this is exactly the equation derived in the finite geometry~\eqref{eq:ShortcutFiniteJoint}. The formula~\eqref{eq:UniversalShortcut} therefore holds in any geometry. The terms that differ between Eq.~\eqref{eq:PsiJointCumulFinal} for the infinite case and Eq.~\eqref{eq:ShortcutFiniteJointScaled} for the finite case come from the fact that in the infinite case, several parameters of the generating functions depend on time $T$, because the dynamics has been coarse grained with the scaling $\Lambda = \sqrt{T}$. These terms do not arise in the finite geometry because the scaling is $\Lambda = L$, which does not depend on time.

\section{Applications}
\label{sec:Applications}

Having obtained a general formula for the shortcut~\eqref{eq:UniversalShortcut}, as well as various specific cases in Section~\ref{sec:Extensions}, we now illustrate the efficiency of this shortcut on different examples.

\subsection{Variance of the generalised current the semi-infinite SEP}

As a first illustration, we consider a semi-infinite simple exclusion process on the positive line, initially at mean density $\bar\rho$, connected at $x=0$ to a reservoir at density $\rho_{L}$. This situation can be described within the MFT formalism, by setting $D(\rho) = 1$ and $\sigma(\rho) = 2 \rho(1-\rho)$ and $\nu=0$ (no drive). As an observable, we consider the generalised current~\eqref{eq:DefGenIntegCurrent0}, which is expressed in terms of the macroscopic density $\rho(x,t)$ by~\eqref{eq:DefGenIntegCurrentMacroFinal}. We again denote $\xi_0$ the point where the current is measured at $t=0$ and $\xi$ the measurement point at $t=1$ (in terms of the macroscopic time $t \in [0,1]$). In the case of the current through the origin ($\xi = \xi_0 = 0$), all the cumulants of the current have been determined~\cite{Grabsch:2024d,Sharma:2024}. Here, we will consider the general case $\xi \neq 0$ and $\xi_0 \neq 0$ and obtain the variance from the shortcut, which takes the form~\eqref{eq:PsiGenCumulQFinal} in this case.

The MFT equations are given by~\cite{Saha:2023}
\begin{align}
    \label{eq:MFTbulkSemiInf}
    \partial_t q
    &= \partial_x^2 q - \partial_x [2q(1-q) \partial_x p]
    \:,
    &
    \partial_t p &= - \partial_x^2 p - (1-2q) (\partial_x p)^2
    \:,
    \\
    \label{eq:MFTinitfinSemiInf}
    q(x,0) &= \lambda \Theta \left(x - \xi_0 \right) + \mu(q(x,0)) - \mu(\bar\rho)
    \:,
    &
    p(x,1) &=  \lambda \Theta \left(x - \xi \right)
    \:,
    \\
    \label{eq:MFTboundSemiInf}
    q(0,t) &= \rho_{L}
    \:,
    &
    p(0,t) &= 0
    \:,
\end{align}
where the chemical potential takes the form
\begin{equation}
    \mu(\rho) = - \ln \left( \frac{1}{r} - 1 \right)
    \:.
\end{equation}
To obtain the variance of the integrated current from the shortcut~\eqref{eq:PsiGenCumulQFinal}, we only need to determine $q(x,1)$ at order $1$ in $\lambda$. We expand $p$ and $q$ as
\begin{equation}
\label{eq:expansion_pq}
    p(x,t) = \lambda p_1(x,t) + O(\lambda^2)
    \:,
    \quad
    q(x,t) = q_0(x,t) +  \lambda q_1(x,t) + O(\lambda^2)
    \:.
\end{equation}
From the MFT equations~(\ref{eq:MFTbulkSemiInf}-\ref{eq:MFTinitfinSemiInf}), we obtain
\begin{equation}
    q_0(x,t) = \rho_{L} \erfc \left( \frac{x}{2 \sqrt{t} } \right)
    + \bar\rho \erf \left( \frac{x}{2 \sqrt{t} } \right)
    \:,
\end{equation}
\begin{equation}
    p_1(x,t) = \frac{1}{2} \erfc \left( \frac{\xi - x}{2 \sqrt{1-t} } \right)
    - \frac{1}{2} \erfc \left( \frac{\xi + x}{2 \sqrt{1-t} } \right)
    \:.
\end{equation}
Making the change of function
\begin{equation}
    q_1 = q_0(1-q_0) p_1 + \tilde{q}_1
    \:,
\end{equation}
we obtain that $\tilde{q}_1$ obeys
\begin{equation}
    \partial_t \tilde{q}_1 = \partial_x^2 \tilde{q}_1
    + (\bar\rho - \rho_L)^2
    \frac{\e^{- \frac{x^2}{2t}}}{\pi t}
    \left[
    \erfc \left( \frac{x + \xi}{2 \sqrt{1-t}} \right)
    - 
    \erfc \left( \frac{\xi - x}{2 \sqrt{1-t}} \right)
    \right]
    \:,
\end{equation}
\begin{equation}
    \tilde{q}_1(x,0) = - \bar\rho (1-\bar\rho) \Theta(x - \xi_0)
    \:,
    \quad
    \tilde{q}_1(0,t) = 0
    \:.
\end{equation}
We have not been able to solve the equations for $\tilde{q}_1$ at all times, but the expression at final time can be computed from the integral representation
\begin{multline}
    \tilde{q}_1(x,1) = (\bar\rho - \rho_L)^2
    \int_0^\infty \dd y \int_0^1 \dd t \:
    G_{1-t}(x|y) \frac{\e^{- \frac{y^2}{2t}}}{\pi t}
    \left[
    \erfc \left( \frac{y + \xi}{2 \sqrt{1-t}} \right)
    - 
    \erfc \left( \frac{\xi - y}{2 \sqrt{1-t}} \right)
    \right]
    \\
    - \frac{1}{2}\bar\rho (1-\bar\rho)
    \left[ 
    \erf \left( \frac{x + \xi_0}{2} \right)
    + 
    \erf \left( \frac{x - \xi_0}{2} \right)
    \right]
    \:,
\end{multline}
where $G_t(x|y)$ is the Green's function that vanish at the origin,
\begin{equation}
    G_t(x|y) = \frac{\e^{- \frac{(x-y)^2}{4t}}}{\sqrt{4\pi t}}
    - \frac{\e^{- \frac{(x+y)^2}{4t}}}{\sqrt{4\pi t}}
    \:.
\end{equation}
The spatial integral can be computed using the table in Ref.~\cite{Owen:1980}. To compute the temporal integral, it is easier to differentiate with respect to $\xi$, to obtain
\begin{equation}
    \partial_\xi \tilde{q}_1(x,1) = 
     (\bar\rho - \rho_L)^2 \left[
     \frac{\e^{- \frac{(x-\xi)^2}{8}}}{\sqrt{2\pi}} 
     \erfc \left( \frac{x + \xi}{2 \sqrt{2}}\right)
     - 
     \frac{\e^{- \frac{(x+\xi)^2}{8}}}{\sqrt{2\pi}} 
     \erfc \left( \frac{|x - \xi|}{2 \sqrt{2}}\right)
     \right]
     \:.
\end{equation}
Integrating over $\xi$ using again~\cite{Owen:1980}, we finally get
\begin{multline}
    \label{eq:q1SemiInfp}
    q_1(x>\xi,1) = - \frac{1}{2}\bar\rho (1-\bar\rho)
    \left[ 
    \erf \left( \frac{x + \xi_0}{2} \right)
    + 
    \erf \left( \frac{x - \xi_0}{2} \right)
    \right]
    + q_0(x,1)(1- q_0(x,1))
    - (\bar\rho - \rho_L)^2 \Bigg[ 2 \erf \left( \frac{x}{2} \right)
    \\
    + \erfc \left( \frac{x}{2} \right) \erfc \left( -\frac{x}{2} \right)
    + \erfc \left( \frac{x - \xi}{2 \sqrt{2}} \right)
    - \erfc \left( -\frac{x + \xi}{2 \sqrt{2}} \right)
    - 4 \mathrm{T} \left( \frac{x-\xi}{2} ,\frac{x + \xi}{x-\xi} \right)
    - 4 \mathrm{T} \left( \frac{x+\xi}{2} ,\frac{x - \xi}{x+\xi} \right)
    \Bigg]
    \:,
\end{multline}
\begin{multline}
    \label{eq:q1SemiInfm}
    q_1(x<\xi,1) = - \frac{1}{2}\bar\rho (1-\bar\rho)
    \left[ 
    \erf \left( \frac{x + \xi_0}{2} \right)
    + 
    \erf \left( \frac{x - \xi_0}{2} \right)
    \right]
    + (\bar\rho - \rho_L)^2 \Bigg[ 2 \erfc \left( \frac{x}{2} \right)
    - \erfc \left( \frac{x - \xi}{2 \sqrt{2}} \right)
    \\
    + \erfc \left( -\frac{x + \xi}{2 \sqrt{2}} \right)
    + 4 \mathrm{T} \left( \frac{x-\xi}{2} ,\frac{x + \xi}{x-\xi} \right)
    - 4 \mathrm{T} \left( \frac{x+\xi}{2} ,\frac{x - \xi}{x+\xi} \right)
    + 8  \mathrm{T} \left( \frac{x}{\sqrt{2}} ,\frac{\xi}{x} \right)
    \Bigg]
    \:,
\end{multline}
where $\mathrm{T}(h,a)$ is Owen's T-function~\cite{Owen:1980}. Similarly, we directly have from the initial condition~\eqref{eq:MFTinitfinSemiInf},
\begin{equation}
    q_1(x,0) = \frac{\bar\rho(1-\bar\rho)}{2}
    \left[
    \erfc \left( \frac{\xi - x}{2 } \right)
    - \erfc \left( \frac{\xi + x}{2} \right)
    - 2 \Theta(x - \xi)
    \right]
    \:.
\end{equation}

In principle, from these expressions one can compute the variance using
\begin{equation}
    \label{eq:ShortcutZero}
    \dt{\hat\psi}{\lambda} = - \bar\rho(\xi - \xi_0)
    + \int_0^\infty [q(\xi + x,1) - q(\xi_0+x,0)] \dd x
    \:,
\end{equation}
but that requires computing the complicated spatial integrals involving~\eqref{eq:q1SemiInfp}. Alternatively, we use here the shortcut~\eqref{eq:PsiGenCumulQFinal}, which yields the cumulant generating function $\hat\psi$ in terms of $q_1(\xi^\pm,1)$ and $\partial_x q_1(\xi^\pm,1)$, which can be easily computed from~(\ref{eq:q1SemiInfp},\ref{eq:q1SemiInfm}). This gives
\begin{equation}
    \hat\psi = \lambda \left( 
    \xi  \erfc\left(\frac{\xi }{2}\right) \left(\bar\rho -\rho _L\right)
    +\frac{2e^{-\frac{\xi ^2}{4}} \left(\rho _L-\bar\rho \right)}{\sqrt{\pi }}+\left(\xi _0-\xi
   \right) \bar\rho
    \right)
    + O(\lambda^2)
    \:.
\end{equation}
This is rather disappointing since it gives only the first cumulant, which can be obtained from $q_0$ using~\eqref{eq:ShortcutZero}. Expanding to next order in $\lambda$, we see that $\hat\psi$ involves a term in $q_2(\xi^+,1) - q_2(\xi^-,1)$, which has not been determined. However, we have argued in Section~\ref{sec:ShortcutGenCurrent} that $r = p - \mu(q)$ is a smooth function at $t=1$. This implies that the boundary condition~\eqref{eq:BoundRelMFT} can be generalised to
\begin{equation}
    \mu(q(\xi^+,1)) - \mu(q(\xi^-,1)) = \lambda
    \:,
\end{equation}
which was first derived in~\cite{Grabsch:2024}. From this relation, we can compute $q_2(\xi^+,1) - q_2(\xi^-,1)$ using our results on $q_1(x,1)$~(\ref{eq:q1SemiInfp},\ref{eq:q1SemiInfm}). Inserting this result into the shortcut~\eqref{eq:PsiGenCumulQFinal} finally gives
\begin{multline}
    \hat\psi = \lambda \left( 
    \xi  \erfc\left(\frac{\xi }{2}\right) \left(\bar\rho -\rho _L\right)
    +\frac{2e^{-\frac{\xi ^2}{4}} \left(\rho _L-\bar\rho \right)}{\sqrt{\pi }}+\left(\xi _0-\xi
   \right) \bar\rho
    \right)
    + \frac{\lambda^2}{2} \Bigg(
    -\xi _0 \left(1-\bar{\rho }\right) \bar{\rho } \erf\left(\frac{\xi -\xi _0}{2}\right)
    \\
    +\erfc\left(\frac{\xi }{2}\right) \left(\bar{\rho }-\rho _L\right) \left(\frac{4 e^{-\frac{\xi ^2}{4}} \left(\bar{\rho }-\rho
   _L\right)}{\sqrt{\pi }}
   -2 \xi  \bar{\rho }
   +\xi \right)
   -\xi  \erfc\left(\frac{\xi }{2}\right)^2 \left(\bar{\rho }-\rho _L\right){}^2
   -2 \left(\sqrt{\frac{2}{\pi }}-\xi \right)
   \erfc\left(\frac{\xi }{\sqrt{2}}\right) \left(\bar{\rho }-\rho _L\right){}^2
   \\
   -\bar{\rho } \left(1-\bar{\rho }\right)
    \left(\xi  \erfc\left(\frac{\xi -\xi_0}{2}\right)
    +\left(\xi +\xi _0\right) \erfc\left(\frac{\xi +\xi_0}{2} \right)\right)
    +\xi 
    \bar{\rho }\left(1-\bar{\rho }\right)
    -2 \sqrt{\frac{2}{\pi }} e^{-\frac{\xi ^2}{2}} \left(\bar{\rho }-\rho _L\right){}^2
    \\
    + \frac{2 e^{-\frac{\xi ^2}{4}} \left(2 \bar{\rho }-1\right) \left(\bar{\rho }-\rho
   _L\right)}{\sqrt{\pi }}
   + 2 \bar{\rho } \left(1-\bar{\rho }\right)
    \frac{e^{-\frac{1}{4} \left(\xi -\xi _0\right){}^2}}{\sqrt{\pi }}
   + 2\bar{\rho } \left(1-\bar{\rho }\right)
   \frac{e^{-\frac{1}{4} \left(\xi +\xi _0\right){}^2}}{\sqrt{\pi }}
   \Bigg)
   + O(\lambda^3)
   \:.
\end{multline}
This gives the expression of the first two cumulants of the generalised integrated current in a semi-infinite SEP. Note that for $\xi_0 = \xi = 0$, we recover the known expression for the current through the origin~\cite{Saha:2023,Grabsch:2024d,Sharma:2024}. In the case $\xi = \xi_0 \to \infty$, we obtain
\begin{equation}
    \hat\psi = \frac{\lambda^2}{2} \frac{\bar\rho(1-\bar\rho)}{\sqrt{\pi}}
    + O(\lambda^3)
    \:,
\end{equation}
which is the result for an infinite system. This is expected since in this case the observable no longer feels the reservoir, so it is as if the system was infinite.

This example illustrates the efficiency of the shortcut presented here: (i) it simplifies the derivation of the cumulant generating function from the solution $q(x,t)$ of the MFT equations and (ii) it even simplifies the first step of solving the MFT equations, as one can directly focus on the solution at the final time, which is often simpler to obtain, as illustrated here.

\subsection{Occupation times in generic systems}

As a second illustration of the shortcut~\eqref{eq:PsiJointCumulFinal} we study the local time $R_T(x)$, defined by~\eqref{eq:localtime0}. We consider an infinite system with no drive $\nu = 0$, initially at equilibrium at a mean constant density $\bar\rho$. We will consider the joint distribution of the local time at the origin (with no loss of generality since the problem is invariant under translations) and at a position $x = \zeta \sqrt{T}$. The joint cumulant generating function takes the form
\begin{equation}
    \label{eq:JointLocTimesExample}
    \moy{
    \moy{\e^{\chi_1 R_T(0) + \chi_2 R_T(x)}}_{\mathrm{e}} }_{\mathrm{i}}
    \underset{T \to \infty}{\simeq}
    \sqrt{T} \:
    \hat\psi \left( \frac{\chi_1}{\sqrt{T}}, \frac{\chi_2}{\sqrt{T}} ; \zeta \right)
    \equiv \sum_{n=1}^\infty \sum_{m=1}^\infty
    \frac{\chi_1^n}{n!} \frac{\chi_2^m}{m!}
    \moy{ R_T(0)^n R_T(x)^m }_c
    \:.
\end{equation}
Note that the scaling form~\eqref{eq:JointLocTimesExample} implies that the cross cumulants have the scaling
\begin{equation}
    \label{eq:DefCrossCumulScaled}
    \moy{R_T(0)^n R_T(x)^m}_c
    \underset{T \to \infty}{\simeq} 
    T^{(1-n-m)/2} \: \kappa_{n,m}(\zeta)
    \:,
    \qquad
    \kappa_{n,m}(\zeta) \equiv
    \partial_{\chi_1}^n \partial_{\chi_2}^m \hat\psi(\chi_1,\chi_2 ; \zeta) \Big|_{(0,0)}
    \:.
\end{equation}
The variance of the occupation time has been computed in~\cite{Meerson:2024} for any $D(\rho)$ and $\sigma(\rho)$. Our goal is to use the shortcut~\eqref{eq:PsiJointCumulFinal} to go beyond and obtain (i) the third cumulant (which measures the skewness of the distribution) $\moy{R_T(0)^3}_c$ and (ii) the covariance $\moy{ R_T(0) R_T(x) }_c$.

\subsubsection{Skewness of the occupation time}

We first study the skewness of the distribution of the local time, encoded in the third cumulant $\moy{R_T(0)^3}_c$. To obtain it from the shortcut~\eqref{eq:PsiJointCumulFinal}, or actually from~\eqref{eq:ShortcutLocalTime} since we consider here only one local time, we only need to determine $q(x=0,t=1)$ at order $2$ in $\chi_1$. As we will see below, this constitutes an important simplification compared to the original MFT procedure used for instance in~\cite{Meerson:2024}. For simplicity, we introduce the scaled cumulant of $R_T(0)$ only, defined from~\eqref{eq:DefCrossCumulScaled} as
\begin{equation}
    \hat\kappa_n \equiv \kappa_{n,0}(\zeta)
    \:,
\end{equation}
which do not depend on $\zeta$ because we set $\chi_2 = 0$.

We again introduce the function $r = p - \mu(q)$, so that the MFT equations~(\ref{eq:MFTLocalTime},\ref{eq:MFTinitfinLocalTime}) take a more symmetric form:
\begin{equation}
\label{eq:MFT_bulk_qr_local}
    \partial_t r 
    = D(q) \partial_x^2 r - \frac{\sigma'(q)}{2} (\partial_x r)^2 - \chi_1 \delta(x)
    \:,
    \hspace{1cm}
    \partial_t p = - D(q) \partial_x^2 p - \frac{\sigma'(q)}{2} (\partial_x p)^2 - \chi_1 \delta(x)
    \:,
\end{equation}
\begin{equation}
    \label{eq:MFT_bound_qr_local}
    r(x,0) = - \mu( \bar\rho)
    \:,
    \hspace{4cm}
    p(x,1) = 0
    \:.
\end{equation}
Our strategy is the following:
\begin{enumerate}[label={\textbf{\arabic*)}},nosep,leftmargin=*]
    \item solve the MFT equations~(\ref{eq:MFT_bulk_qr_local},\ref{eq:MFT_bound_qr_local}) for $(q,r)$ at order 1 in $\chi_1$;
    \item determine $r$ at order $2$ in $\chi_1$ only at the origin and at final time to deduce $q(0,1) = \mu^{-1} \big( -r(0,1) \big)$ at order $2$ in $\chi_1$;
   \item Use the shortcut~\eqref{eq:ShortcutLocalTime} to obtain $\hat\kappa_3$.
\end{enumerate}
We thus expand $p$, $q$ and $r$ at order 2 in $\chi_1$
\begin{align}
    r = - \mu(\bar\rho) +  \chi_1 r_1 + \chi_1^{2} r_2+ O(\chi_1^3)
    \:,
    \quad
    p = \chi_1 p_1 + \chi_1^{2} p_2+ O(\chi_1^3)
    \:,
    \quad
    q = \bar\rho + \chi_1 q_1 + \chi_1^{2} q_2+ O(\chi_1^3)
    \:,
\end{align}
with
\begin{equation}
\label{eq:qfromr}
    q_1 \big|_{t=1} = - \frac{\sigma(\bar\rho)}{2D(\bar\rho)} r_1 \Bigg|_{t=1}
    \:,
    \quad
    q_2 \big|_{t=1}
    = \frac{\sigma(\bar\rho) \sigma'(\bar\rho)}{8D(\bar\rho)^2} r_1^2
    - \frac{D'(\bar\rho) \sigma(\bar\rho)^2}{8D(\bar\rho)^3} r_1^2
    - \frac{\sigma(\bar\rho)}{2D(\bar\rho)} r_2
    \Bigg|_{t=1}
    \:,
\end{equation}
due to the definition of $r$.

At first order in $\chi$, the equations~(\ref{eq:MFT_bulk_qr_local},\ref{eq:MFT_bound_qr_local}) become:
\begin{equation}
    \label{eq:MFTbulkR1}
    \partial_t r_1 
    = D(\bar\rho) \partial_x^2 r_1  -  \delta(x)\:,
    \hspace{1cm}
    \partial_t p_1 = - D(\bar\rho) \partial_x^2 p_1 -  \delta(x)
    \:,
\end{equation}
\begin{equation}
    \label{eq:MFTiniR1}
    r_1(x,0) = 0
    \:, \hspace{3cm}
    p_1(x,1) = 0
    \:.
\end{equation}
Which can be solved straigthforwardly for $r_1$:
\begin{equation}
    r_1(x,t) = -\frac{\sqrt{t}}{\sqrt{\pi D(\bar\rho)}}
    \e^{-\frac{x^2}{4 t D(\bar\rho)}} 
    +\frac{x}{2 D(\bar\rho)}
    \erfc\left(\frac{x}{2 \sqrt{t D(\bar\rho)}}\right)
    -\frac{x \: \Theta (-x)}{D(\bar\rho)}
     \:.
\end{equation}
From~\eqref{eq:qfromr} we obtain $q_1$,
\begin{equation}
    q_1(0,1) = -\frac{\sigma(\bar\rho)}{2D(\bar\rho)}r_1(0,1)
\end{equation}
Using the shortcut relation~\eqref{eq:ShortcutLocalTime} we directly get:
\begin{equation}
    \label{eq:kappa2R}
    \hat{\kappa}_2=\frac{2 \sigma (\bar\rho )}{3 \sqrt{\pi } D(\bar\rho)^{3/2}}
    \:,
\end{equation}
which coincides with the expression previously obtained in~\cite{Meerson:2024} by solving the MFT equations. Note that to determine only $\hat{\kappa}_2$ we do not need to solve the full MFT equations, rather only obtain $r_1$ at the origin, at final time.

The second MFT field $p_1$ is obtained by noticing that the substitution $\tilde{t}=1-t$ implies that $p_1(x,\tilde{t})$ obeys the same equation as $r_1$ replacing $\chi_1$ by $-\chi_1$. Thus, 
\begin{equation}
    p_1(x,t) = -r_1(x,1-t)
     \:.
\end{equation}

At second order in $\chi_1$, we only need to determine $r_2(x=0,t=1)$ to obtain $q_2$ from~\eqref{eq:qfromr}, since we have already determined $r_1$.
Expanding~\eqref{eq:MFT_bulk_qr_local} at second order in $\chi_1$, we get
\begin{equation}
    \label{eq:Rr2}
    \partial_t r_2 
    = D(\bar\rho) \partial_x^2 r_2 
    + D'(\bar\rho) q_1 \partial_x^2 r_1
    -  \frac{\sigma'(\bar\rho)}{2}\left(\partial_x r_1\right)^{2}\:, \hspace{1cm} r_2(x,0) = 0\:.
\end{equation}
This diffusion equation can be solved formally as
\begin{equation}
\label{eq:formallyr2}
    r_2(z,t)=\int_{-\infty}^{\infty}\dd x \int_{0}^{t} \dd s \: K_{t-s}(z-x)
    \left[
    D'(\bar\rho) q_1 \partial_x^2 r_1
    -  \frac{\sigma'(\bar\rho)}{2}\left(\partial_x r_1\right)^{2}
    \right]
    \:,
\end{equation}
where $K_{t}(x) =\frac{\e^{-\frac{x^{2}}{4D(\bar\rho) t}}}{\sqrt{4\pi D(\bar\rho) t}}$ is the Gaussian kernel.
The key point is that the expression~\eqref{eq:formallyr2} drastically simplifies when setting $z=0$ and $t=1$, to give
\begin{align}
    r_2(0,1)&=\int_{-\infty}^{\infty}\dd x \int_{0}^{1} \dd s \: K_{1-s}(x) \left[
    D'(\bar\rho) q_1 \partial_x^2 r_1
    -  \frac{\sigma'(\bar\rho)}{2}\left(\partial_x r_1\right)^{2}
    \right]
    \nonumber
    \\
    & =\frac{(4+\pi ) \sigma (\bar\rho ) D'(\bar\rho)-2 (\pi -2) D(\bar\rho) \sigma '(\bar\rho )}{16 \pi  D(\bar\rho)^3}
    \:,
\end{align}
where we have made use of~\cite{Owen:1980} to compute the spatial integrals.
From this result, we obtain $q_2(x=0,t=1)$ using~\eqref{eq:qfromr}, and thus the third cumulant from the shortcut~\eqref{eq:ShortcutLocalTime},
\begin{equation}
    \hat{\kappa}_3=
    \frac{3 \sigma (\bar\rho ) }{32 \pi  D(\bar\rho )^4}
    \Big( 
    2 \pi  D(\bar\rho ) \sigma '(\bar\rho )
    -(\pi+8) \sigma (\bar\rho ) D'(\bar\rho )\Big)
    \:,
\end{equation}
illustrating the efficiency of this approach.

\subsubsection{Covariance of occupation times}

We now turn to the study of the covariance between two local times $\moy{R_T(0) R_T(x=\zeta \sqrt{T})}_c$.
We again introduce the function $r = p - \mu(q)$, so that $p$ and $r$ obey
\begin{equation}
    \partial_t r 
    = D(q) \partial_x^2 r - \frac{\sigma'(q)}{2} (\partial_x r)^2 - \chi_1 \delta(x) - \chi_2 \delta(x-\zeta)
    \:,
    \hspace{1cm}
    \partial_t p = - D(q) \partial_x^2 p - \frac{\sigma'(q)}{2} (\partial_x p)^2 - \chi_1 \delta(x) - \chi_2 \delta(x-\zeta)
    \:,
\end{equation}
\begin{equation}
    r(x,0) = - \mu( \bar\rho)
    \:,
    \hspace{4cm}
    p(x,1) = 0
    \:.
\end{equation}
Solving the equation for $r$ at first order in $\chi_1$ and $\chi_2$ and using that $q(x,1) = \mu^{-1}\big( -r(x,1) \big)$ because $p$ vanishes at final time yields
\begin{align}
    q(x=0,t=1)
    &=
    \bar\rho + 
    \frac{\sigma(\bar\rho)}{2 D(\bar\rho)} \left[
    \frac{1}{\sqrt{\pi D(\bar\rho)}} 
    \left(
    \chi_1 
    + \chi_2 \e^{- \frac{\zeta^2}{4D(\bar\rho)}}
    \right)
    - \frac{\chi_2 |\zeta|}{2 D(\bar\rho)} 
    \erfc \left( \frac{|\zeta|}{\sqrt{4 D(\bar\rho)}} \right)
    \right]
    + \cdots
    \:,
    \\
    q(x=\zeta,t=1)
    &=
    \bar\rho + 
    \frac{\sigma(\bar\rho)}{2 D(\bar\rho)} \left[
    \frac{1}{\sqrt{\pi D(\bar\rho)}} 
    \left(
    \chi_1 \e^{- \frac{\zeta^2}{4D(\bar\rho)}}
    + \chi_2
    \right)
    - \frac{\chi_1 |\zeta|}{2 D(\bar\rho)} 
    \erfc \left( \frac{|\zeta|}{\sqrt{4 D(\bar\rho)}} \right)
    \right]
    + \cdots
    \:.
\end{align}
From the shortcut~\eqref{eq:PsiJointCumulFinal}, the scaled cumulant generating function~\eqref{eq:JointLocTimesExample} obeys
\begin{equation}
    \hat\psi + \chi_1 \partial_{\chi_1} \hat\psi + \chi_2 \partial_{\chi_2} \hat\psi - \zeta \partial_\zeta \hat\psi =
    2 \chi_1 q(0,1) + 2 \chi_2 q(\zeta,1)
    \:.
\end{equation}
Inserting the above expressions of $q(0,1)$ and $q(\zeta,1)$, taking a derivative with respect to $\chi_1$ and $\chi_2$ and evaluating at $\chi_1 = \chi_2 = 0$ gives a differential equation for the scaled covariance $\kappa_{1,1}$ defined in~\eqref{eq:DefCrossCumulScaled}
\begin{equation}
    - \zeta \kappa_{1,1}'(\zeta) + 3 \kappa_{1,1}(\zeta)
    = \frac{\sigma(\bar\rho)}{D(\bar\rho)}
    \left[
    \frac{2}{\sqrt{\pi D(\bar\rho)}} \e^{- \frac{\zeta^2}{4D(\bar\rho)}}
    - \frac{|\zeta|}{D(\bar\rho)} \erfc \left( \frac{|\zeta|}{\sqrt{4 D(\bar\rho)}} \right)
    \right]
    \:.
\end{equation}
Solving this differential equation yields
\begin{equation}
    \kappa_{1,1}(\zeta)
    = \frac{\sigma(\bar\rho)}{2 D(\bar\rho)}
    \left[
    \left(
    \frac{\zeta^2}{4 D(\bar\rho)} + 1
    \right)
    \frac{\e^{-\frac{\zeta^2}{4 D(\bar\rho)}}}{3 \sqrt{\pi D(\bar\rho)}}
    - \frac{\zeta}{6 D(\bar\rho)} \left(
    \frac{\zeta^2}{D(\bar\rho)} + 6
    \right)
    \erfc \left( \frac{|\zeta|}{\sqrt{4 D(\bar\rho)}} \right)
    \right]
    + c |\zeta|^3
    \:,
\end{equation}
with an integration constant $c$. The constant can be determined by imposing that $\kappa_{1,1}(\zeta \to \infty) = 0$ so that the two local times become uncorrelated when they are infinitely far apart. This gives $c=0$ and thus
\begin{equation}
    \moy{R_T(0) R_T(\zeta \sqrt{T})}_c 
    \underset{T \to \infty}{\simeq}
    \frac{\sigma(\bar\rho)}{2 D(\bar\rho) \sqrt{T}}
    \left[
    \left(
    \frac{\zeta^2}{4 D(\bar\rho)} + 1
    \right)
    \frac{\e^{-\frac{\zeta^2}{4 D(\bar\rho)}}}{3 \sqrt{\pi D(\bar\rho)}}
    - \frac{\zeta}{6 D(\bar\rho)} \left(
    \frac{\zeta^2}{D(\bar\rho)} + 6
    \right)
    \erfc \left( \frac{|\zeta|}{\sqrt{4 D(\bar\rho)}} \right)
    \right]
    \:.
\end{equation}
This result again illustrates the efficiency of the shortcut~\eqref{eq:PsiJointCumulFinal} to compute the joint statistical properties of various observables.

\section{Conclusion}

We have proved a general shortcut relation~\eqref{eq:UniversalShortcut} to obtain the joint cumulants of different observables, both for finite and infinite systems.
This relation extends the ones recently conjectured for the case of the integrated current through the origin~\cite{Grabsch:2024b,Berlioz:2025}.
We have first derived the shortcut relation in Section~\ref{sec:shortcutQt} by relying on two conservation laws that result from the Hamiltonian structure of the MFT equations. We have derived several extensions in Section~\ref{sec:Extensions} using an alternative approach, in particular in situations in which the conservation laws of Section~\ref{sec:shortcutQt} are broken. The breaking of these conservation relations introduces additional derivatives with respect to parameters of the cumulant generating function, as in Eq.~\eqref{eq:ShortcutLocalTime}. See Appendix~\ref{app:genobs} for more details. Nevertheless, for all the observables considered in this article, the cumulant generating function is fully determined by the solution of the MFT equations at the final time only. This surprising feature greatly simplifies the evaluation of the cumulant generating function, as shown in Section~\ref{sec:Applications}.

We have considered observables that are local in space: the integrated current through a given point, and the local time at a specific point. Our analysis extends to nonlocal observables, such as the local time in a given interval, as shown in Appendix~\ref{app:genobs}. In this case, the shortcut relation~\eqref{eq:ShortcutLocalInt} still involves a spatial integral over the MFT profile at the final time. This is due to the specific choice of the observable, which by construction involves integrating over different points in space. Nevertheless, the shortcut relation remains a huge simplification compared to the original expression in terms of the action that involves an integral over both space and time.

We have illustrated on three examples (the variance of a generalised current in the semi-infinite SEP, the skewness of the local time in any infinite system and the covariance of local times in the same situation) that this approach is of practical use and indeed simplifies a lot the computations.

We stress that, in the case of the integrated current, this shortcut is actually a generalisation of Fick's law which gives not only the mean current but all the cumulants, and thus the full distribution. In this context, the MFT profile at final time $q(x,T)$ plays the role of a generalised mean density, which contains all the correlations between the density and the integrated current, as discussed below Eq.~\eqref{eq:DefGenProf}. This further underlines the important physical role played by this generalised profile~\cite{Poncet:2021,Grabsch:2022,venturelli2024,Berlioz:2025,Grabsch:2025b}.

Finally, we focused on the one dimensional situation since it is the most studied case in the MFT context. The extension of the shortcut~\eqref{eq:UniversalShortcut} to higher dimensional systems is still open.

\section*{Acknowledgments}

A. Grabsch thanks Thibault Congy for stimulating discussions, in particular concerning the transformation~\eqref{eq:ChangeVarR}.

\appendix 

\section{Alternative computation of the integrals at $t=1$ or $t=0$}
\label{sec:AppRegulInteg}

We provide here an alternative computation of the integrals in Eq.~\eqref{eq:CumulNew3}, which are ill-defined since $\partial_x p(x,1) = \lambda \delta(x)$ and $q$ is discontinuous at the origin at $t=1$. We thus regularise this integral as
\begin{equation}
    \int_{-\infty}^\infty \sigma(q) \partial_x p \Big|_{t=1} \dd x
    = \lim_{\varepsilon \to 0}
     \int_{-\infty}^\infty \sigma(q) \partial_x p \Big|_{t=1-\varepsilon} \dd x
     \:.
\end{equation}
We need to determine the behaviour of $p$ and $q$ near $x=0$ for $t$ close to $1$.  Close to the discontinuities, the MFT equations~\eqref{eq:MFTbulk} are expected to be dominated by the diffusion, therefore we look for solutions near $x=0$ in the form
\begin{equation}
    \label{eq:ScalingFormsT1}
    p(x,t) = P \left( \frac{x}{\sqrt{1-t}} \right)
    \:,
    \quad
    q(x,t) = Q \left( \frac{x}{\sqrt{1-t}} \right)
    \:,
    \quad \text{with} \quad
    P( \pm \infty) = p(0^\pm,1)
    \:,
    \quad
    Q( \pm \infty) = q(0^\pm,1)
    \:.
\end{equation}
Inserting these scaling forms into the MFT equations~\eqref{eq:MFTbulk} and replacing $x \to z \sqrt{1-t}$ yields
\begin{equation}
    z P' + 2 D(Q) P''(Q) + \sigma'(Q) (P')^2 = 0
    \:,
    \quad
    z Q' - 2 \partial_z (D(Q) Q' - \sigma(Q) P') = 0
    \:.
\end{equation}
Isolating $z$ in the first equation, inserting it in the second one and multiplying the result by $P'$, we obtain
\begin{equation}
    \partial_z \left(
    2 D(Q) Q' P' - \sigma(Q) (P')^2
    \right)
    = 0
    \:.
\end{equation}
Since both $P$ and $Q$ tend to constants at infinity, see Eq.~\eqref{eq:ScalingFormsT1}, we have that
\begin{equation}
    \label{eq:RelScalingFroms}
    2 D(Q) P' Q' - \sigma(Q) (P')^2
     = 0
    \:.
\end{equation}
With the scaling forms~\eqref{eq:ScalingFormsT1}, we can compute
\begin{equation}
    \lim_{\varepsilon \to 0}
     \int_{-\infty}^\infty \sigma(q) \partial_x p \Big|_{t=1-\varepsilon} \dd x
     = \int_{-\infty}^\infty
     \sigma(Q(z)) P'(z)
     \dd z
     = 2 \int_{-\infty}^\infty
     D(Q(z)) Q'(z)
     \dd z
     = 2 \int_{Q(-\infty)}^{Q(+\infty)} D(r) \dd r
     \:.
\end{equation}
Using the values at infinity of $Q$~\eqref{eq:ScalingFormsT1} and inserting the result into~\eqref{eq:CumulNew3}, we obtain
\begin{equation}
    \int_{-\infty}^\infty \sigma(q) \partial_x p \Big|_{t=1} \dd x
    = 2 \int_{q(0^-,1)}^{q(0^+,1)} D(r) \dd r
    \:,
\end{equation}
which coincides with~\eqref{eq:IntegT1}, as it should.

\section{More general observables from the conservation relations}
\label{app:genobs}

Let us consider a more general observable of the form
\begin{equation}
    \label{eq:DefGenObs}
    \mathcal{O}[\rho] = \int_0^1 \dd t \int_{-\infty}^\infty \dd x \: f(x,t) G[\rho(x,t)]
    \:,
\end{equation}
for an infinite one dimensional system. For $f(x,t) = \delta(x-\zeta)$, this reduces to the local time at position $\zeta$. The scaled cumulant generating function~\eqref{eq:CumulQtFinal} now takes the form
\begin{equation}
    \label{eq:DefCumulGenObs}
    \hat\psi = -S[q,p] - F[q(x,0)] + \lambda \mathcal{O}[q]
    \:,
\end{equation}
where $S$ is given by~\eqref{eq:ActionHamilton} and $F$ by~\eqref{eq:DistInit}.
The last term breaks the invariance of the action under spatial and temporal translations, at the origin of the conservation relations~\eqref{eq:ConsH} and~\eqref{eq:Consqdxp} that were essential in the derivation of the shortcut in Section~\ref{sec:shortcutQt}. We discuss in this Appendix how the cumulant generating function~\eqref{eq:DefCumulGenObs} can nevertheless be simplified in this case.

The MFT equations associated to~\eqref{eq:DefGenObs} can be derived by minimising the action. They now read (we consider the case $\nu=0$ for simplicity)
\begin{equation}
    \label{eq:MFTgen}
  \partial_t q 
  =
  \partial_{x} \left[D(q)\partial_{x}q- \sigma(q)\partial_{x}p
  \right]
  \:,
  \quad
    \partial_{t}p =
    -D(q)\partial_{x}^{2}p
    -\frac{1}{2}\sigma'(q)(\partial_{x}p)^{2}
    - \lambda f G'[q]
    \:,
\end{equation}
with the initial and final conditions
\begin{equation}
\label{eq:MFTinitfinGen}
    p(x,1) = 0
    \:,
    \quad
    p(x,0) = \int_{\bar\rho}^{q(x,0)} \frac{2 D(r)}{\sigma(r)} \dd r
    \:.
\end{equation}
Although the conservation relations~\eqref{eq:ConsH} and~\eqref{eq:Consqdxp} are now explicitly broken, we can show from the MFT equations~\eqref{eq:MFTgen} that
\begin{equation}
\label{eq:ConsqdxpExt}
    \partial_t (q \partial_x p)
    + 
    \partial_x \left[
    \frac{\sigma(q)}{2} (\partial_x p)^2 - D(q) \partial_x q \partial_x p
    - q \partial_t p
    \right]
    = \lambda f G'[q] \partial_x q
    \:.
\end{equation}
\begin{equation}
    \label{eq:ConsHgen}
    \partial_t \left[ 
    \frac{\sigma(q)}{2} (\partial_x p)^2 - D(q) \partial_x q \partial_x p
    \right] +
    \partial_x \left[
    D(q) \partial_x p \partial_x q
    + (D(q) \partial_x q - \sigma(q)\partial_x p ) \partial_t p
    \right]
    = - \lambda f G'[q] \partial_t q
    \:.
\end{equation}
Proceeding as in Section~\ref{sec:shortcutQt}, we obtain from~\eqref{eq:ConsqdxpExt}
\begin{equation}
    S[q,p] = -2 \int_0^1 \dd t \int_{-\infty}^\infty \dd x \left[
    \frac{\sigma(q)}{2} (\partial_x p)^2 - D(q) \partial_x q \partial_x p
    \right]
    + \int_{-\infty}^\infty \big[ x q \partial_x p + q p \big]_{t=0}^{t=1} \dd x
    - \lambda \int_0^1 \dd t \int_{-\infty}^{\infty} \dd x \: x f G'[q] \partial_x q
    \:.
\end{equation}
While from~\eqref{eq:ConsHgen} we get
\begin{equation}
    \int_0^1 \dd t \int_{-\infty}^\infty \dd x \left[
    \frac{\sigma(q)}{2} (\partial_x p)^2 - D(q) \partial_x q \partial_x p
    \right]
    =  \left[
    \frac{\sigma(q)}{2} (\partial_x p)^2 - D(q) \partial_x q \partial_x p
    \right] \bigg|_{t=1}
    + \lambda \int_0^1 \dd t \int_{-\infty}^\infty \dd x \: t f G'[q] \partial_t q
    \:.
\end{equation}
Using these expressions in the cumulant generating function~\eqref{eq:DefCumulGenObs}, we get
\begin{multline}
    \hat\psi = 2 \int_{-\infty}^\infty \dd x
    \left[
    \frac{\sigma(q)}{2} (\partial_x p)^2 - D(q) \partial_x q \partial_x p
    \right] \bigg|_{t=1}
    - \int_{-\infty}^\infty [x q \partial_x p + q p]_{t=0}^{t=1} 
    - F[q(x,0)]
    \\
    + \lambda \int_0^1 \dd t \int_{-\infty}^\infty \dd x [x f G'[q] \partial_x q + 2 t f G'[q] \partial_t q + f G[q]]
    \:.
\end{multline}
Inserting the initial and final condition~\eqref{eq:MFTinitfinGen} together with an integration by parts, we get for an initially flat density profile $\bar\rho$,
\begin{equation}
    \hat\psi =  \lambda \int_0^1 \dd t \int_{-\infty}^\infty \dd x [f G'[q] \partial_x q + 2 t f G'[q] \partial_t q + f G[q]]
    \:.
\end{equation}
After an integration by parts in time, we can write this expression as
\begin{equation}
    \hat\psi =  \lambda \int_0^1 \dd t \int_{-\infty}^\infty \dd x 
    \left[x f \partial_x G[q] - 2 t G[q]  \partial_t f-  f G[q] \right]
    + 2 \lambda \int_{-\infty}^\infty f(x,1) G[q(x,1)] \dd x
    \:.
\end{equation}
Using that, since $(q,p)$ is the minimum of the action,
\begin{equation}
    \partial_\lambda \hat\psi = \mathcal{O}[q] = \int_0^1 \dd t \int_{-\infty}^\infty \dd x f G[q]
    \:,
\end{equation}
we obtain
\begin{equation}
    \label{eq:RelPsiGenInterm}
    \hat\psi + \lambda \partial_\lambda \hat\psi 
    =  \lambda \int_0^1 \dd t \int_{-\infty}^\infty \dd x  
    \left[x f \partial_x G[q] - 2 t G[q] \partial_t f  \right]
    + 2 \lambda \int_{-\infty}^\infty f(x,1) G[q(x,1)] \dd x
    \:.
\end{equation}
The last term involves only the MFT profile at the final time, while the other terms involve the full time evolution of the profile $q(x,t)$. Nevertheless, for specific observables, this term can be simplified to yield a shortcut that involves only $q(x,1)$. Let us discuss a few cases.
\begin{itemize}
    \item As in the case of the local time, choosing $f(x,t) = \delta(x-\zeta)$, with $\zeta$ fixed, Eq.~\eqref{eq:RelPsiGenInterm} simplifies as
    \begin{equation}
    \hat\psi + \lambda \partial_\lambda \hat\psi 
    =  \lambda \zeta \int_0^1 \dd t \: \partial_x G[q]\big|_{x=\zeta}
    + 2 \lambda G[q(\zeta,1)] \dd x
    \:.
    \end{equation}
Using again that $(q,p)$ minimise the action, we have
\begin{equation}
    \partial_\zeta \hat\psi = \lambda \int_0^1 \dd t \: \partial_x G[q] \Big|_{x= \xi} 
    \:,
\end{equation}
so that
    \begin{equation}
    \label{eq:PsiShortcutGenLocal}
    \boxed{
    \hat\psi + \lambda \partial_\lambda \hat\psi - \zeta \partial_\zeta \hat\psi
    =
    2 \lambda G[q(\zeta,1)] \dd x
    \:,
    }
    \end{equation}
    which indeed reduces to~\eqref{eq:ShortcutLocalTime} in the case of the local time $G[q] = q$. Note that the shortcut~\eqref{eq:PsiShortcutGenLocal} applies to more general observables, such as the activity studied in several works~\cite{Rolland:2008,Vanicat:2021} which corresponds to $G[q] = \sigma(q)$.
    \item We can consider more general observables inside an interval $[\zeta,\zeta']$. Choosing $f(x,t) = \un_{[\zeta,\zeta']}(x)$, with $\un_{[\zeta,\zeta']}(x) = 1$ if $x \in [\zeta,\zeta']$ and $0$ otherwise, we obtain from~\eqref{eq:RelPsiGenInterm}
    \begin{equation}
    \hat\psi + \lambda \partial_\lambda \hat\psi 
    =  \lambda \int_0^1 \dd t   
    \left[\zeta' G[q] \big|_{x=\zeta'} - \zeta G[q] \big|_{x=\zeta}  \right]
    -  \lambda \int_0^1 \dd t  \int_{\zeta}^{\zeta'} G[q]
    + 2 \lambda \int_{\zeta}^{\zeta'} G[q(x,1)] \dd x
    \:.
\end{equation}
Using again that $(q,p)$ minimise the action, we have
\begin{equation}
    \partial_{\zeta'} \hat\psi = \lambda \int_0^1 \dd t \: G[q(\zeta',t)]
    \:,
    \quad
    \partial_{\zeta} \hat\psi = -\lambda \int_0^1 \dd t \: G[q(\zeta,t)]
    \:,
    \quad
    \partial_\lambda \hat\psi = \int_0^1 \dd t  \int_{\zeta}^{\zeta'} G[q]
    \:.
\end{equation}
Hence, we finally obtain
    \begin{equation}
    \label{eq:ShortcutLocalInt}
    \boxed{
    \hat\psi + 2 \lambda \partial_\lambda \hat\psi 
    - \zeta' \partial_{\zeta'} \hat\psi
    + \zeta \partial_\zeta \hat\psi
    = 2 \lambda \int_{\zeta}^{\zeta'} G[q(x,1)] \dd x
    \:,
    }
    \end{equation}
    which again only involves the MFT profile at the final time. Note that there remains a spatial integral at the final time, due to the choice of the observable which is itself an integral.
\end{itemize}
We have seen on these examples that it is possible to simplify the shortcut~\eqref{eq:RelPsiGenInterm} so that it involves the MFT profile at the final time only, but at the cost of introducing a derivative with respect to the parameters of $f$. This is actually a general feature. Indeed, assuming that $G[q]$ decays to infinity, we can write~\eqref{eq:RelPsiGenInterm} as
\begin{equation}
    \label{eq:RelPsiGenInterm}
    \hat\psi + 2 \lambda \partial_\lambda \hat\psi 
    =  -\lambda \int_0^1 \dd t \int_{-\infty}^\infty \dd x  
    \left[x \partial_x f + 2 t \partial_t f  \right] G[q] 
    + 2 \lambda \int_{-\infty}^\infty f(x,1) G[q(x,1)] \dd x
    \:.
\end{equation}
Using now that $(q,p)$ minimise the action, we get
\begin{equation}
    \frac{\delta \hat\psi}{\delta f(x,t)} = \lambda G[q(x,t)]
    \:,
\end{equation}
so that
\begin{equation}
    \label{eq:RelPsiGenInterm}
    \boxed{
    \hat\psi + 2 \lambda \partial_\lambda \hat\psi 
    +  \int_0^1 \dd t \int_{-\infty}^\infty \dd x  
    \left[x \partial_x f + 2 t \partial_t f  \right] \frac{\delta \hat\psi}{\delta f} 
    =  2 \lambda \int_{-\infty}^\infty f(x,1) G[q(x,1)] \dd x
    \:.
    }
\end{equation}
Formally, this expression shows that the cumulant generating function $\hat\psi$ is fully determined by the MFT profile at the final time $q(x,1)$. However, in practice it is more convenient to start from Eq.~\eqref{eq:RelPsiGenInterm} and then simplify it depending on the specific choice of $f$.

Finally, note that the derivation presented in this Appendix can in principle also be performed in the case of a finite system discussed in Section~\ref{sec:finite}. However, in this case one gets additional terms due to the boundaries of the spatial domain of integration, coming for instance from the spatial integration of the second term in~\eqref{eq:ConsHgen}. Such terms involve $\partial_x q$ or $\partial_x p$ at the boundaries in space. These are not easily determined from the MFT equations~\eqref{eq:MFTjointFinite} and~\eqref{eq:BC-space}. Therefore, the derivation of the shortcut~\eqref{eq:ShortcutFiniteJointScaled} from the method presented in this Appendix is still an open question.

\section{Alternative derivation of the action}
\label{sec:AppHomogeneousHamilt}

In this Appendix, we present an alternative derivation of the
expression~(\ref{eq:ActionNew2}) of the action that relies on the form
of the Hamiltonian density~(\ref{eq:ActionHamilton}). The component
$T^{11}$ of the stress-energy tensor~(\ref{eq:DefStressEnerTensor}) is
defined in terms of the Lagrangian
$\mathcal{L} = p \partial_t q - \mathcal{H}$ by
\begin{equation}
  T^{11} = \dep{\mathcal{L}}{(\partial_x q)} \partial_x q
  +  \dep{\mathcal{L}}{(\partial_x p)} \partial_x p
  - \mathcal{L}
  \:.
\end{equation}
Since the Hamiltonian density~(\ref{eq:ActionHamilton}) is a
homogeneous function of order $2$ in $\partial_x q$ and $\partial_x p$
(it only involves terms of the form
$(\partial_x q)^n(\partial_x p)^{2-n}$ with $n=0,1$ or $2$), we get
that
\begin{equation}
  T^{11} = - 2 \mathcal{H} - \mathcal{L}
  \:.
\end{equation}
Hence, the conservation relation~(\ref{eq:ConsT01}) can be written as
\begin{equation}
  \partial_t T^{01} = \partial_x (2 \mathcal{H} + \mathcal{L})
  \:.
\end{equation}
Multiplying by $x$ and integrating by parts yields
\begin{equation}
  \partial_t \int_{-\infty}^\infty x T^{01} \dd x
  = - 2 \int_{-\infty}^\infty \mathcal{H} \dd x
  - \int_{-\infty}^\infty \mathcal{L} \dd x
  \:.
\end{equation}
Integrating over time and using that the Hamiltonian is conserved, see
Eq.~(\ref{eq:ConsHamilt}), we get
\begin{equation}
  \int_{-\infty}^\infty \left[ x T^{01} \right]_{t=0}^{t=1} \dd x
  = - 2 \int_{-\infty}^\infty \mathcal{H} \big|_{t=1} \dd x
  - S[q,p]
  \:.
\end{equation}
Using the expression of $T^{01}$~(\ref{eq:DefStressEnerTensor}) and
performing an integration by parts on the l.h.s., we recover the
expression of the action~(\ref{eq:ActionNew2}).

In this alternative derivation, the fact that the Hamiltonian density
is a homogeneous function of order $2$ in $\partial_x q$ and
$\partial_x p$ plays a central role. In general, this structure can
however be broken by additional terms. For instance, in the case of
weakly driven systems considered in Section~\ref{sec:WeakDrive}, the
Hamiltonian density becomes
\begin{equation}
  \mathcal{H}(q,p) = \frac{\sigma(q)}{2} (\partial_x p)^2 - D(q) \partial_x q \partial_x p
  + \nu \sigma(q) \partial_x p
  \:.
\end{equation}
Due to the last term which is linear in $\partial_x p$, the component
$(11)$ of the stress energy tensor becomes
\begin{equation}
  T^{11} = - 2 \mathcal{H} - \mathcal{L} + \nu \sigma(q) \partial_x p
  \:.
\end{equation}
Repeating the steps presented above, we get
\begin{equation}
  \int_{-\infty}^\infty \dd x \int_0^1 \dd t \: \mathcal{L}
  = - 2 \int_{-\infty}^\infty \mathcal{H} \big|_{t=1} \dd x
  - \int_{-\infty}^\infty \left[ x T^{01} \right]_{t=0}^{t=1} \dd x
  + \int_{-\infty}^{\infty} \dd x \int_0^1 \dd t \: \nu \sigma(q) \partial_x p
  \:.
\end{equation}
The last term comes from the fact that the Hamiltonian density is no
longer a homogeneous function of order $2$ in $\partial_x q$ and
$\partial_x p$. It can nevertheless be expressed in terms of the
cumulant generating function $\hat\psi$ as shown in
Section~\ref{sec:WeakDrive}, since
\begin{equation}
  \nu \partial_\nu \hat\psi =
  \int_{-\infty}^{\infty} \dd x \int_0^1 \dd t \: \nu \sigma(q) \partial_x p
  \:.
\end{equation}
Within this approach, the additional $\nu \partial_\nu \hat\psi$ term
present in the shortcut~(\ref{eq:PsiShortcutDrive}) appears as a
consequence of the breaking of the homogeneous structure of the
Hamiltonian density.

Note that this discussion holds for the case of infinite systems
only. Indeed, in the case of finite systems discussed in
Section~\ref{sec:finite}, although the Hamiltonian is not homogenous
in $\partial_x p$ and $\partial_x q$ when $\nu \neq 0$, the additional
$\nu \partial_\nu \hat\psi$ term is not present in the
shortcut~\eqref{eq:ShortcutFiniteJointScaled}. A unified description
of the finite and infinite cases can be obtained by considering the
time derivative of the cumulant generating function, as discussed in
Section~\ref{sec:UniversalShortcut}.

\bibliographystyle{apsrev4-1}

\end{document}